\documentclass[aps,prl,reprint,superscriptaddress]{revtex4-2}

\usepackage{amsthm}
\usepackage{graphicx}
\usepackage{dcolumn}
\usepackage{bm}
\usepackage{overpic}
\newtheorem*{theorem}{Theorem}

\theoremstyle{definition}

\theoremstyle{remark}

\pdfoutput=1
\usepackage[utf8]{inputenc}
\usepackage[T1]{fontenc}

\usepackage{xcolor, soul}
\usepackage[normalem]{ulem}
\usepackage{amsfonts}
\usepackage{amssymb}
\usepackage{amsmath}
\usepackage{verbatim}
\usepackage{multirow}
\usepackage{natbib}

\newcommand{\bk}{\mathbf{k}}
\newcommand{\eqm}{\xi_\text{QM}}
\newcommand{\ef}{\xi_\text{flat}}
\newcommand{\ecoh}{\xi_\text{coh}}
\newcommand{\tmx}{t_\text{max}}

\begin{document}

\preprint{APS/123-QED}

\title{Embedding Independent Length Scale of Flat Bands}

\author{Seokju Lee}
\affiliation{Department of Physics and Astronomy, Seoul National University, Seoul 08826, Korea}
\affiliation{Center for Theoretical Physics (CTP), Seoul National University, Seoul 08826, Korea}
\affiliation{Institute of Applied Physics, Seoul National University, Seoul 08826, Korea}
\author{Seung Hun Lee}
\affiliation{Department of Physics and Astronomy, Seoul National University, Seoul 08826, Korea}
\affiliation{Center for Theoretical Physics (CTP), Seoul National University, Seoul 08826, Korea}
\affiliation{Institute of Applied Physics, Seoul National University, Seoul 08826, Korea}
\author{Bohm-Jung Yang}
\email{bjyang@snu.ac.kr}
\affiliation{Department of Physics and Astronomy, Seoul National University, Seoul 08826, Korea}
\affiliation{Center for Theoretical Physics (CTP), Seoul National University, Seoul 08826, Korea}
\affiliation{Institute of Applied Physics, Seoul National University, Seoul 08826, Korea}

\begin{abstract}
In flat-band systems with quenched kinetic energy, most of the conventional length scales related to the band dispersion become ineffectual. Although a few geometric length scales, such as the quantum metric length, can still be defined, because of their embedding dependence, i.e., the dependence on the choice of orbital positions used to construct the tight‑binding model, they cannot serve as a universal length scale of the flat-band systems.
Here, we introduce an embedding independent length scale $\ef$ of a flat band that is defined as the localization length of an in‑gap state proximate to the flat band. 
Because $\ef$ is derived from the intrinsic localization of compact localized states, it is solely determined by the Hamiltonian and provides a robust foundation for embedding independent observables. We show analytically that the superconducting coherence length in a flat‑band superconductor is given by $\ef$ in the weak‑coupling limit, thereby identifying $\ef$ as the relevant length scale for many‑body phenomena. 
Numerical simulations on various lattice models confirm all theoretical predictions, including the correspondence between $\ef$ and the superconducting coherence length. Our results highlight $\ef$ as a universal length scale for flat bands and open a pathway to embedding independent characterization of interacting flat‑band materials.
\end{abstract}

\maketitle


\textit{Introduction.---}
Flat bands provide a fertile platform for strongly correlated phenomena, as their vanishing kinetic energy amplifies interaction and geometric effects. This unique feature has revealed flat-band magnetism~\cite{Tasaki1992ferromagnetism,Yin2019}, geometry driven superconductivity~\cite{Cao2018unconventional,Tian2023Dirac}, and fractionalization~\cite{Neupert2011fractional,Parameswaran2013} rarely observed in dispersive bands.

In dispersive bands, characteristic lengths such as the coherence length or mean free path are directly tied to the dispersion, reflecting kinetic energy scales. When the bandwidth vanishes, these conventional lengths also vanish or lose their meaning. In flat bands, the only length scale that remains finite is the quantum metric length (QML). In one dimension, it is defined as~\cite{Hu2023}
\begin{align}\label{eq qml}
    \eqm=\sqrt{\int_{-\pi}^{\pi}\frac{dk}{2\pi}g(k)},
\end{align}
where $g(k)$ is the quantum metric,
\begin{align}\label{eq qm}
    g(k)=
    \langle\partial_k u(k)|(1-|u(k)\rangle\langle u(k)|)|\partial_k u(k)\rangle,
\end{align}
and $|u(k)\rangle$ is the periodic part of the Bloch function. The QML has been known to describe the gauge-invariant part of Wannier spreading~\cite{maxwf} and provides the geometric contribution to the superfluid weight~\cite{Peotta2015}.

A key feature of $\eqm$ is its embedding dependence, i.e., dependence on intracell orbital positions (orbital embedding). In tight-binding models, orbital embedding provides additional information beyond the hopping amplitudes~\cite{embed}. Embedding-independent quantities, such as the band structure, depend only on hopping parameters, whereas geometric quantities like the Berry curvature depend on embedding. Since universal relations exist only among quantities sharing the same embedding dependence, identifying this property is essential~\cite{embed}. Although it has recently been reported that $\eqm$ determines the superconducting coherence length in flat bands~\cite{Iskin2023,Chen2024}, the two lengths differ in their embedding dependence, and their equivalence holds only at the level of order of magnitude~\cite{stub}. 
To remedy this issue, the minimal quantum metric, defined as the smallest value of $\eqm$ obtained over all possible embeddings, was introduced and shown to exhibit deep connections to superconductivity~\cite{herzogarbeitman2022manybodysuperconductivity,PhysRevB.106.014518}. However, because the minimal $\eqm$ is also constructed entirely from the quantum metric, searching a distinct and intrinsically embedding-independent length scale remains crucial to properly characterize general flat-band phenomena.

Here, we introduce a universal, embedding-independent flat-band length scale $\ef$. We define $\ef$ as the localization length of an in-gap state induced by a local perturbation, in the limit where its energy approaches the flat band energy. Flat bands host not only Bloch states but also compact localized states (CLSs), which are eigenstates confined to a few unit cells. When a local perturbation creates an in-gap state near the flat band energy, its response propagates through the overlap between CLSs, and this propagation length is set by $\ef$. We show that $\ef$ is determined solely by CLS overlaps, and is therefore embedding-independent. Furthermore, in the weak-coupling limit of a flat-band superconductor, the superconducting coherence length $\ecoh$ coincides with $\ef$. This identifies $\ef$ as the fundamental length scale governing intrinsic flat-band properties.

\textit{Embedding dependence of $\eqm$.---}
An embedding-independent quantity remains invariant under shifts of intracell orbital positions while keeping the real-space tight-binding Hamiltonian fixed, as in the case of the band structure~\cite{embed}. In contrast, embedding-dependent quantities vary explicitly with the choice of orbital embedding. 

To illustrate the embedding dependence of $\eqm$, consider a translation-symmetric tight-binding Hamiltonian $H$ with a real-space basis $|R,\alpha\rangle$, denoting an orbital $\alpha$ at position $\tau_\alpha$ within the unit cell at $R$. Here, we ignore other internal degrees of freedoms such as spin for convenience. In momentum space, the corresponding basis can be written in two distinct forms:
\begin{align}
|k,\alpha\rangle_N &= \frac{1}{N_\text{cell}}\sum_R e^{ik(R+\tau_\alpha)}|R,\alpha\rangle, \\
|k,\alpha\rangle_P &= \frac{1}{N_\text{cell}}\sum_R e^{ikR}|R,\alpha\rangle,
\end{align}
where the periodic basis $|k,\alpha\rangle_P$ satisfies $|k+G,\alpha\rangle_P=|k,\alpha\rangle_P$ for any reciprocal vector $G$, while $|k,\alpha\rangle_N$ is non-periodic.

In these two bases, $H$ takes the forms
\begin{align}\label{eq blochh}
H &= \sum_{k\alpha\beta}|k,\alpha\rangle_N H_N(k)_{\alpha\beta}\langle k,\beta|_N \nonumber \\
  &= \sum_{k\alpha\beta}|k,\alpha\rangle_P H_P(k)_{\alpha\beta}\langle k,\beta|_P.
\end{align}
The matrices $H_N(k)$ and $H_P(k)$ share the same eigenvalue $E_n(k)$ but have different eigenvectors: the non-periodic $u_{n,N}(k)$ and the periodic $u_{n,P}(k)$, respectively. The Bloch state is
\begin{align}
|\psi_n(k)\rangle=\sum_{\alpha}u_{n,\eta}(k)_{\alpha}|k,\alpha\rangle_{\eta},
\end{align}
where $\eta=N,P$, and the two eigenvectors are related by
\begin{align}\label{eq utaum}
u_{n,N}(k)_{\alpha}=e^{-ik\tau_\alpha}u_{n,P}(k)_{\alpha}
\end{align}
up to a U(1) gauge. Since $H_P(k)$ is defined in the periodic basis $|k,\alpha\rangle_P$, it is embedding-independent, and so is $u_{n,P}(k)$. In contrast, $u_{n,N}(k)$ inherits embedding dependence through Eq.~\eqref{eq utaum}.

The Bloch vector $|u(k)\rangle$ used in Eq.~\eqref{eq qm} corresponds to $|u_{n,N}(k)\rangle$. Under a shift $\tau_\alpha \to \tau_\alpha + \delta\tau_\alpha$, $u_{n,N}(k)$ changes by a phase factor and the corresponding quantum metric $g_n(k)$ acquires terms linear in $\delta\tau_\alpha$. It confirms that $g_n(k)$, and hence its integral $\eqm$, depend explicitly on the orbital embedding (see Supplemental Material (SM) for details).

\begin{figure}
\begin{overpic}[width=0.29\linewidth]{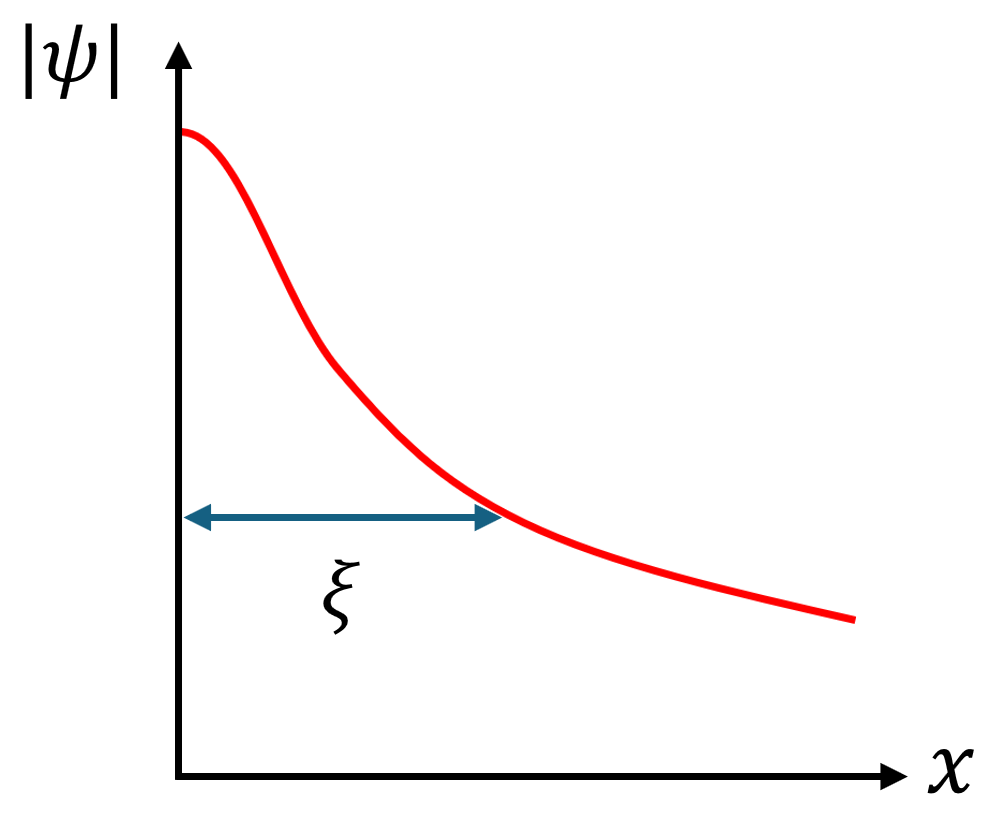}
\put(-2,90){\text{(a)}}
\end{overpic}
\begin{overpic}[width=0.34\linewidth]{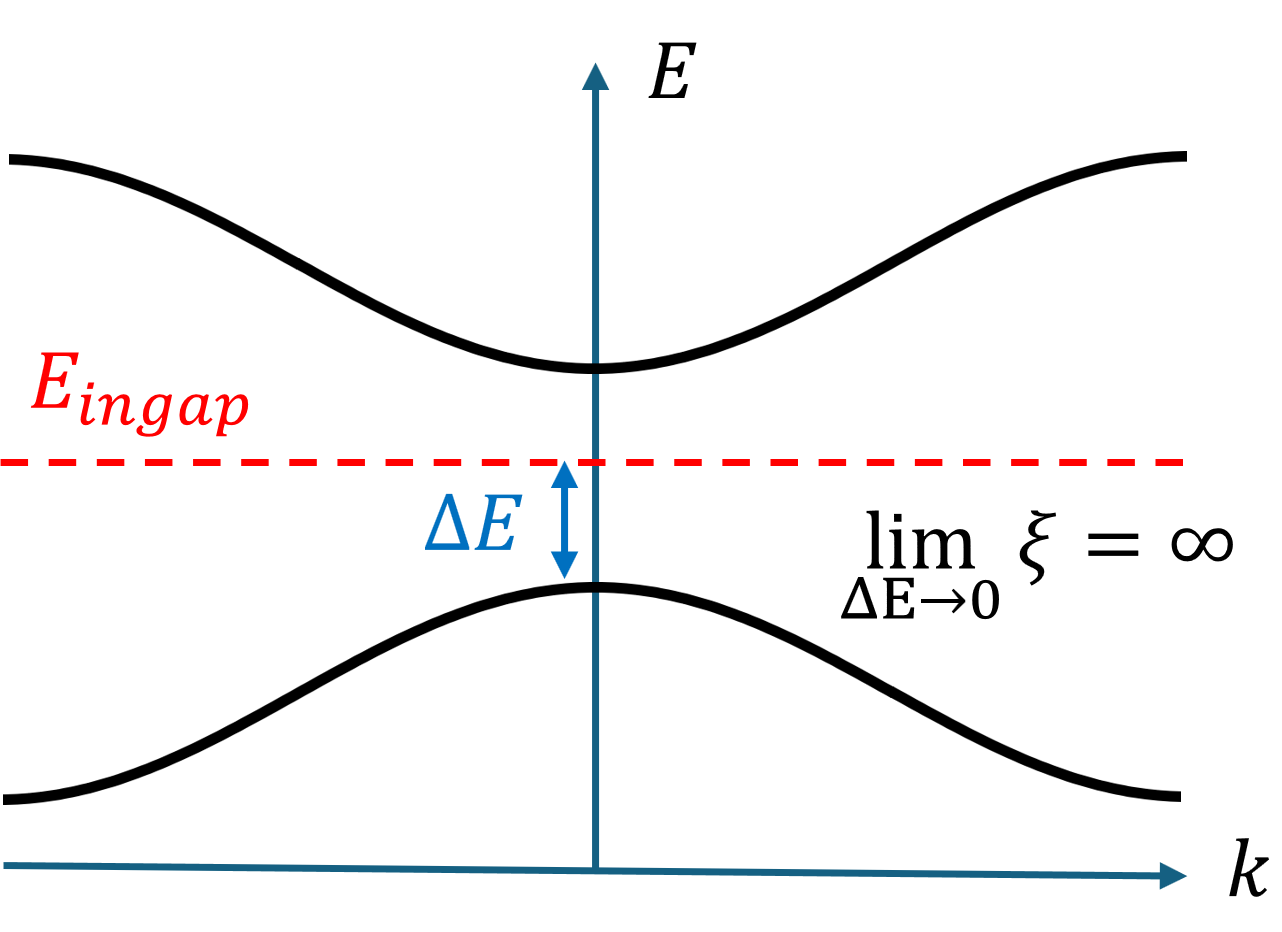}
\put(-2,76){\text{(b)}}
\end{overpic}
\begin{overpic}[width=0.34\linewidth]{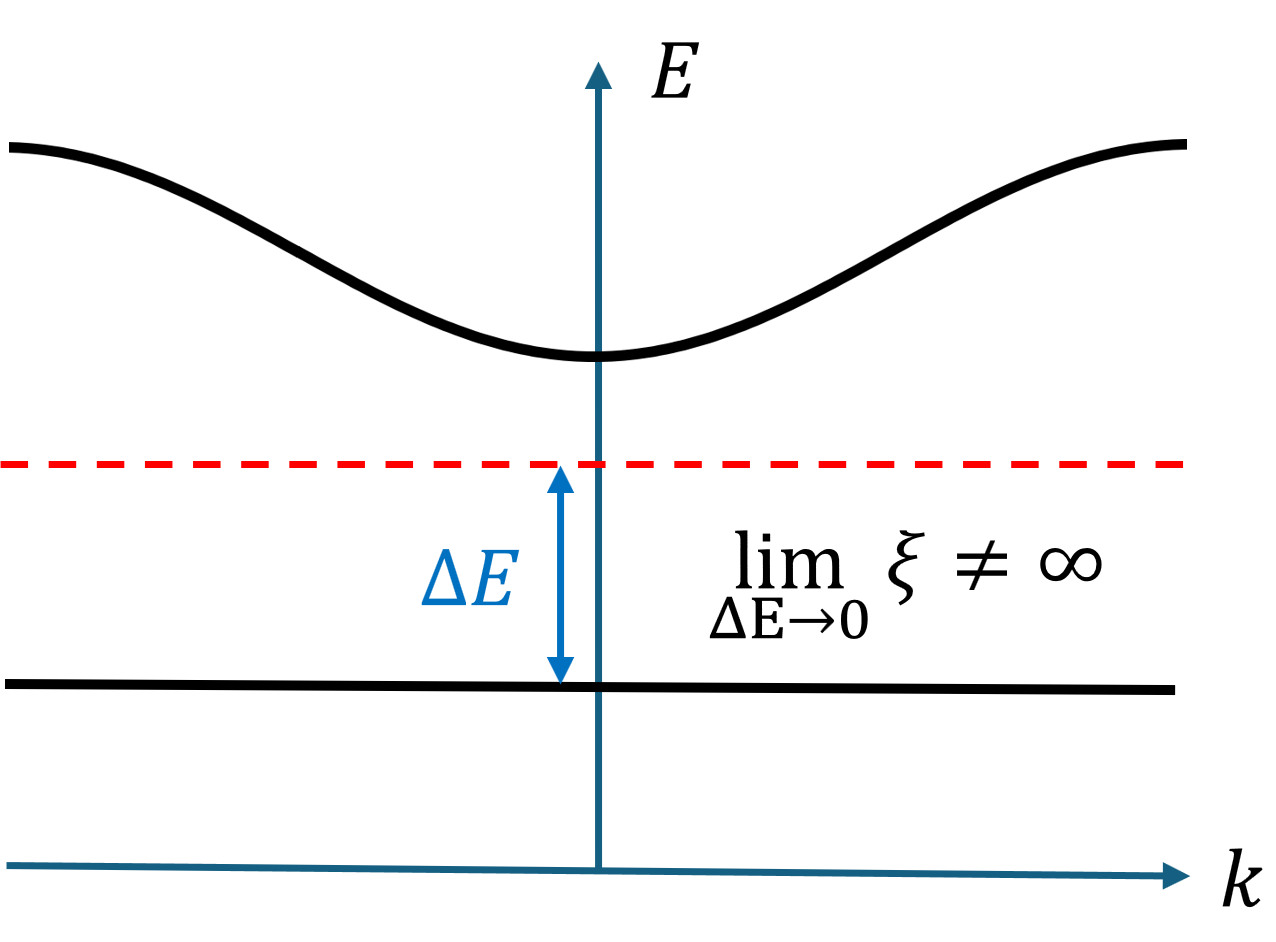}
\put(-2,76){\text{(c)}}
\end{overpic}
\caption{\label{fig:boundary}(a) Wave function profile of an exponentially localized in-gap state with localization length $\xi$. (b) A gapped band structure with an in-gap state. As the in-gap state energy (dashed red) approaches a dispersive bulk band (black), $\xi$ diverges. (c) When the in-gap state approaches a flat band, $\xi$ remains finite.
}
\end{figure}

\textit{Localization length of in-gap states.---}
We next analyze in-gap states induced by local perturbations and show that their localization is controlled by the decay of the bare Green’s function, which leads to an embedding-independent length $\ef$.

Consider a one-dimensional, noninteracting tight-binding Hamiltonian
\begin{align}
H_{0}=\sum_{k,n}E_{n}(k)\,c_{nk}^\dagger c_{nk},
\end{align}
with Bloch momentum $k$, band index $n$, and band energy $E_n(k)$. In the unit cell–orbital basis $(i\alpha)$, the retarded Green’s function is
\begin{align}\label{eq:green0}
G^{0}(i\alpha,j\beta;\omega)
=\frac{1}{N}\sum_{n,k}
\frac{u_{n,\alpha}(k)\,u^{*}_{n,\beta}(k)}
     {\omega+i\eta-E_{n}(k)}\,e^{ik(R_i-R_j)},
\end{align}
where $N$ is the number of unit cells and $u_n(k)$ is the periodic Bloch eigenvector. (Non-periodic Bloch vector gives only additional phase $e^{ik(\tau_\alpha-\tau_\beta)}$ that does not affect the decay.)

Let $V$ be a local perturbation, i.e., supported on a finite number of unit cells. An in-gap bound state $|\psi_b\rangle$ at energy $\omega_b$ satisfies
\begin{align}
(H_0+V)|\psi_b\rangle=\omega_b|\psi_b\rangle,
\end{align}
which can be written in Lippmann–Schwinger form,
\begin{align}\label{eq:LS}
|\psi_b\rangle=G^{0}(\omega_b)\,V\,|\psi_b\rangle.
\end{align}
Projecting onto the real-space orbital basis, with $\psi_\alpha(R_i)=\langle i\alpha|\psi_b\rangle$, gives
\begin{align}\label{eq:LS_real}
\psi_\alpha(R_i)
= \sum_{j\beta} G^{0}(i\alpha,j\beta;\omega_b)\,\langle j\beta|V|\psi_b\rangle.
\end{align}
Since $V$ is local, $\langle j\beta|V|\psi_b\rangle$ is nonzero only within a finite set of unit cells near the perturbation. Thus the asymptotic decay of $\psi_\alpha(R_i)$ at large separations $r=R_i-R_j$ is governed entirely by the spatial decay of the Green’s function $G^{0}(i\alpha,j\beta;\omega_b)$ (see SM for details).

In particular, for a pointlike perturbation $V = V_0\,c_{i_0\alpha_0}^\dagger c_{i_0\alpha_0}$, the in-gap state at energy $\omega_b$ has a wavefunction equivalent to the Green's function up to an overall normalization constant. (A detailed derivation is given in the SM.) We therefore define it as
\begin{align}\label{eq:psi0_def}
\psi^0_\alpha(R_i-R_{i_0})
\;\equiv\;
G^{0}(i\alpha,i_0\alpha_0;\omega_b).
\end{align}
We will use $\psi^0_\alpha$ as a representative in-gap wavefunction to analyze the large-distance decay.

\textit{Asymptotic decay in dispersive and flat bands.---}
If $\omega_b$ approaches the edge of a dispersive band $E_n(k)$, the dominant contributions to Eq.~\eqref{eq:green0} come from a finite set of momenta $\{k_j\}$ minimizing $|\omega_b-E_n(k)|$. The bound state then reduces to a superposition of plane waves $e^{ik_j r}$, and the localization length diverges [see Fig.~\ref{fig:boundary}(a,b)].

In contrast, when $\omega_b$ approaches an isolated flat band with $E_n(k)=E_{\text{flat}}$, the denominator in Eq.~\eqref{eq:green0} is $k$-independent and all momenta contribute evenly. Defining $|\varepsilon|\equiv|E_{\text{flat}}-\omega_b|\ll 1$ and retaining only the flat-band term, the in-gap wavefunction $\psi^0_\alpha(R_i)$ takes the form,
\begin{align}\label{eq:flat_proj_short}
\psi^0_\alpha(r)\approx\frac{1}{N}\sum_k
\frac{u_\alpha(k)u_{\alpha_0}^*(k)}{\varepsilon}\,e^{ikr},
\qquad r=R_i-R_{i_0},
\end{align}
which is the Fourier transform of $u_\alpha(k)u_{\alpha_0}^*(k)$. Treating $k$ as a complex number, the Paley–Wiener theorem~\cite{Paley1934} implies that if $u_\alpha(k)u_{\alpha_0}^*(k)$ is analytic for $|\mathrm{Im}\,k|<\gamma$ and develops its nearest singularity at $|\mathrm{Im}\,k|=\gamma$, then the in-gap wavefunction decays as
\begin{align}\label{eq asymp}
|\psi^0_\alpha(r)|\sim A e^{-\gamma|r|}\quad(|r|\to\infty),
\end{align}
where $A$ is a prefactor. Such exponential decay shown in Fig.~\ref{fig:boundary}(c) is consistent with the recent report of the real-space decay of flat-band projectors in one dimension~\cite{kim2025realspacedecayflat}.

We quantify this decay using the standard definition of localization length,
\begin{align}\label{eq:xi_def_short}
\xi(\psi)\equiv
\Bigl\{
\lim_{|x|\to\infty}
\frac{1}{|x|}\ln\frac{1}{|\psi(x)|}
\Bigr\}^{-1},
\end{align}
so that $|\psi(x)|\propto e^{-|x|/\xi}$ for $x\gg a$, with $a$ the lattice constant. We then define the flat-band localization length
\begin{align}\label{eq:ef_def_short}
\ef
=
\lim_{\omega_b\to E_\text{flat}}\xi(\psi^0_\alpha),
\end{align}
which is therefore finite and, crucially, embedding-independent, since $\psi^0_\alpha$ is embedding-independent as given by Eq.~\eqref{eq:flat_proj_short}. We also note that, as discussed in the SM, the localization length of an in-gap state proximate to a flat band remains finite regardless of its origin. However, its localization length does not necessarily coincide with $\ef$ depending on the details of the system.

\textit{Relation to CLS and localization length.---}
Interestingly, $\ef$ is expressed purely in terms of CLS overlaps. We denote the normalized CLS centered at lattice vector $R$ by $|\mathrm{CLS};R\rangle$. A flat-band Bloch eigenstate with momentum $k$ can then be written as
\begin{align}\label{eq clseig}
    |\psi_{\mathrm{flat}}(k)\rangle
    = \sum_{R} e^{ikR}\,|\mathrm{CLS};R\rangle .
\end{align}
Since the periodic Bloch vector $u_\alpha(k)$ satisfies
\begin{align}
|\psi_{\mathrm{flat}}(k)\rangle \propto \sum_{R,\alpha} e^{ikR}\,u_{\alpha}(k)\,|R,\alpha\rangle,
\end{align}
it follows that $u_\alpha(k)$ can be expressed in terms of the CLS at $R=0$ as
\begin{align}\label{eq cls}
    u_{\alpha}(k)
    = \frac{e^{i\phi(k)}}{N(k)}
      \sum_{R}
      \langle R,\alpha\,|\,\mathrm{CLS};0\rangle\,e^{-ikR},
\end{align}
where $N(k)$ ensures $\|u(k)\|=1$ and $\phi(k)$ is U(1) phase. Defining the CLS overlaps
\begin{align}\label{eq clsoverlap}
    \lambda_{t} \equiv \bigl\langle\mathrm{CLS};R\,\bigl|\,\mathrm{CLS};R-t\bigr\rangle,
    \qquad t\in\mathbb{Z},
\end{align}
the normalization constant $N(k)$ becomes
\begin{align}\label{eq nkm}
    N(k)
    = \sqrt{\sum_{t\in\mathbb{Z}}\lambda_{t}\,e^{ikt}}
    = \sqrt{1 + \lambda_{1}e^{ik} + \lambda_{-1}e^{-ik} + \cdots} .
\end{align}

Now let us determine the analytic region of
\begin{align}\label{eq fourfunc}
    u_{\alpha}(k)\,u^{*}_{\alpha_0}(k) 
    =\frac{1}{|N(k)|^2}
      \Bigl(\sum_{R}
      \langle R,\alpha|\mathrm{CLS};0\rangle\,e^{-ikR}\Bigr)\\
      \times\Bigl(\sum_{R'}
      \langle \mathrm{CLS};0|R',\alpha_0\rangle\,e^{ikR'}\Bigr).\nonumber
\end{align}
For real $k$, this function is analytic because both $|N(k)|^2\,=N(k)^2$ and the Fourier series are analytic. As long as $N(k)\neq 0$, the analyticity extends to $k\in\mathbb{C}$ via analytic continuation. Hence, the boundary of the analytic region is set by the zeros of $N(k)$ (or $N(k)^2$ equivalents) and $\ef$ obeys
\begin{align}\label{eq efconm}
    \ef=\max\Bigl\{\frac{1}{|\mathrm{Im}\,k|}\;:\;\sum_{t\in\mathbb{Z}}\lambda_{t}\,e^{ikt}=0\Bigr\}.
\end{align}
This general result is consistent with the localization length of the flat band projector proposed recently based on models with short CLS overlaps~\cite{DiBenedetto2025}.

\begin{figure}
\raisebox{8mm}{
\begin{overpic}[width=0.49\linewidth]{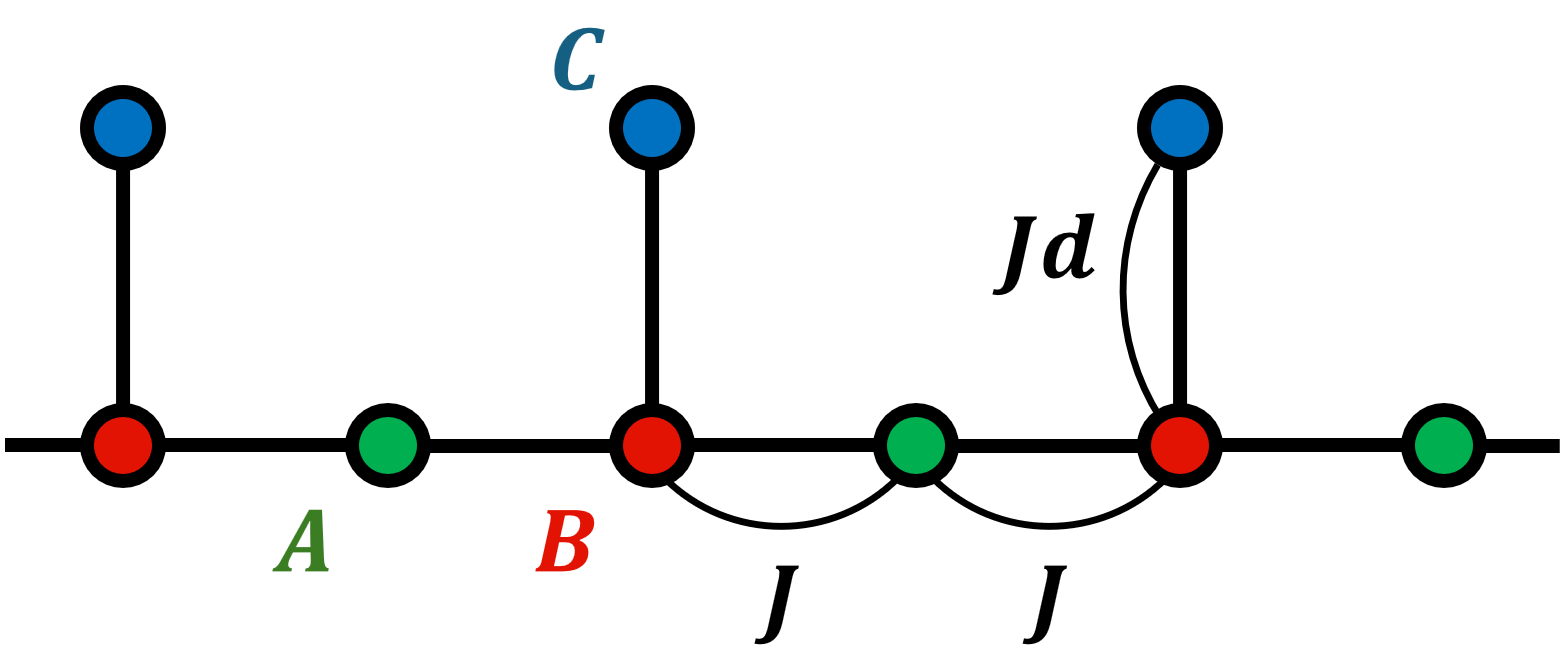}
\put(-2,54){\text{(a)}}
\end{overpic}}
\begin{overpic}[width=0.47\linewidth]{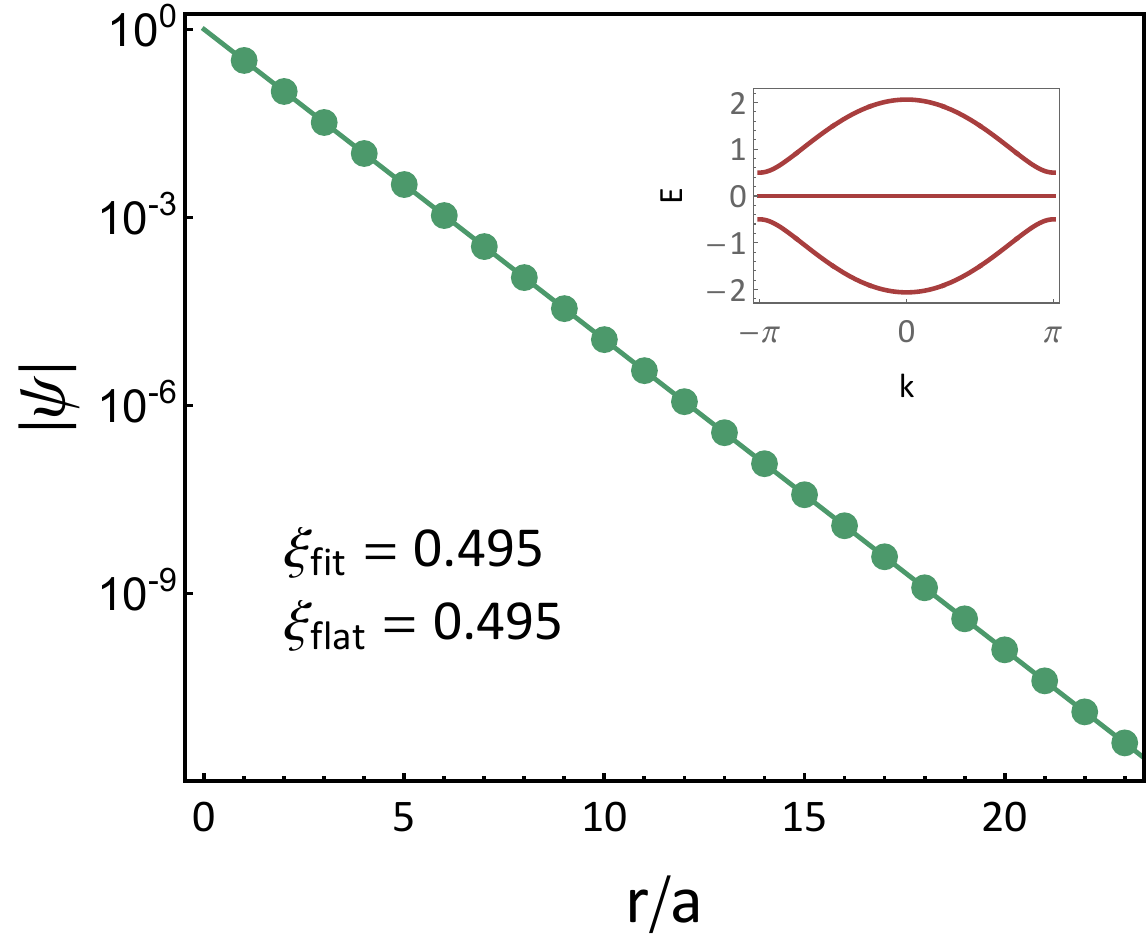}
\put(-4,76){\text{(b)}}
\end{overpic}
\caption{\label{fig:ingap} (a) Stub lattice model with three orbitals ($A,B,C$) per unit cell and hopping amplitudes $J$ and $Jd$. (b) Real-space profile of the in-gap state induced by a point impurity. The localization length extracted from an exponential fit is shown together with the theoretical prediction. The inset shows the band structure. The data are obtained for $J=1$ and $d=0.5$, in which applying a local potential $U=0.01$ to a single $C$ orbital generates an in-gap state at $E=0.0024$.
}
\end{figure}

We illustrate the validity of Eq.~\eqref{eq efconm} by using the one-dimensional Stub lattice model [Fig.~\ref{fig:ingap}(a)] with three orbitals $\{A,B,C\}$ per unit cell~\cite{stub} [More examples are shown in SM]. The tight-binding model with hopping scale $J$ and asymmetry $d$ [see Fig.~\ref{fig:ingap}(a)] hosts one perfectly flat band and two dispersive bands. The CLS and the flat-band Bloch vector in the $\{A,B,C\}$ basis are
\begin{align}\label{eq stubcls}
    |\mathrm{CLS};R\rangle
    =\frac{1}{\sqrt{2+d^2}}
      \bigl\{d\,|R,A\rangle+|R-1,C\rangle+|R,C\rangle\bigr\},
\end{align}
\begin{align}\label{eq stubvec}
    u(k)=\frac{1}{N(k)\sqrt{2+d^2}}
    \begin{bmatrix}
    d\\[2pt]
    0\\[2pt]
    1+e^{ik}
    \end{bmatrix},
\end{align}
which has only nearest-neighbor CLS overlaps other than $\lambda_0=1$ such that
\begin{align}\label{eq stubov}
    \lambda_1=\lambda_{-1}=\frac{1}{2+d^2}.
\end{align}
Thus
\begin{align}\label{eq nk2}
    N(k)=\sqrt{1+2\lambda_1 \cos k}
         =\sqrt{1+\frac{2}{2+d^2}\cos k}.
\end{align}
From Eq.~\eqref{eq efconm} we obtain
\begin{align}\label{eq loclen}
    \ef
    =\Bigl[\operatorname{arccosh}\!\Bigl(\frac{2+d^2}{2}\Bigr)\Bigr]^{-1}
    =\Bigl[\operatorname{arccosh}\!\Bigl(\frac{1}{|2\lambda_{1}|}\Bigr)\Bigr]^{-1}.
\end{align}
Here, the first equality uses the Stub-lattice parameter $d$, whereas the second equality—expressed via the CLS overlap $\lambda_{1}$—is the form that applies to any flat band whose CLS has support on two unit cells with $\lambda_{t}=0$ for $|t|\ge2$, consistent with Ref.~\cite{DiBenedetto2025}. As shown in Fig.~\ref{fig:ingap}(b), an in-gap state induced by a small on-site potential near the flat band remains exponentially localized, and the fitted $\ef$ agrees with the theoretical value.

Since $\ef$ is determined solely by CLS overlaps, two distinct flat-band models with different microscopic Hamiltonians and CLS shapes share the same $\ef$ as long as their overlap set $\{\lambda_t\}$ is identical. Moreover, in-gap states created by weak perturbations near the flat-band energy exhibit exponential decay with the decay length governed primarily by the underlying single-particle structure, suggesting that $\ef$ serves as a unified length scale for localization in flat-band many-body phenomena such as flat-band superconductivity as discussed below.

\textit{Relation to superconducting coherence length.---}
The superconducting coherence length $\ecoh$ is a representative embedding-independent length of a many-body ground state. In BCS theory applied to a dispersive band, one finds $\ecoh=\hbar v_F/\Delta$, confirming that it depends only on embedding-independent quantities—the Fermi velocity $v_F$ and the mean-field superconducting gap $\Delta$. We note that since long-range order is absent in one dimension, the coherence length may not be unambiguously defined in a strict sense; here we work at the mean-field level and take $\ecoh$ to denote the exponential decay length of the anomalous correlator.

To establish $\ecoh=\ef$ for flat-band superconductors in the weak-coupling regime, we consider
\begin{align}\label{eq stubsupm}
    H=\sum_{i,j,\alpha,\beta,\sigma}t_{ij,\alpha\beta}\,c^{\dagger}_{i\alpha,\sigma}c_{j\beta,\sigma}
        -\mu N+U\sum_{i\alpha}n_{i\alpha,\uparrow}n_{i\alpha,\downarrow},
\end{align}
with on-site attraction $U=-|U|<0$ and chemical potential $\mu$ at the flat band energy.

The coherence length $\ecoh$ can be extracted from the anomalous correlator
\begin{align}\label{eq cf}
    K_{\alpha}(R_j-R_i)
    =
    \bigl\langle
        c_{i\alpha,\uparrow}\,c_{j\alpha,\downarrow}
    \bigr\rangle,
\end{align}
which decays exponentially in one dimension~\cite{stub} as,
\begin{align}
    K_{\alpha}(r)\sim e^{-|r|/\ecoh},\qquad r=R_j-R_i\gg a.
\end{align}
Since the real-space eigenstates of Eq.~\eqref{eq stubsupm} are determined by $t_{ij,\alpha\beta}$, $\mu$, and $U$—and not by the choice of intracell orbital positions—the eigenstates, and therefore $K_\alpha(R)$ and $\ecoh$, are embedding-independent.

\begin{figure}
\begin{overpic}[width=0.47\linewidth]{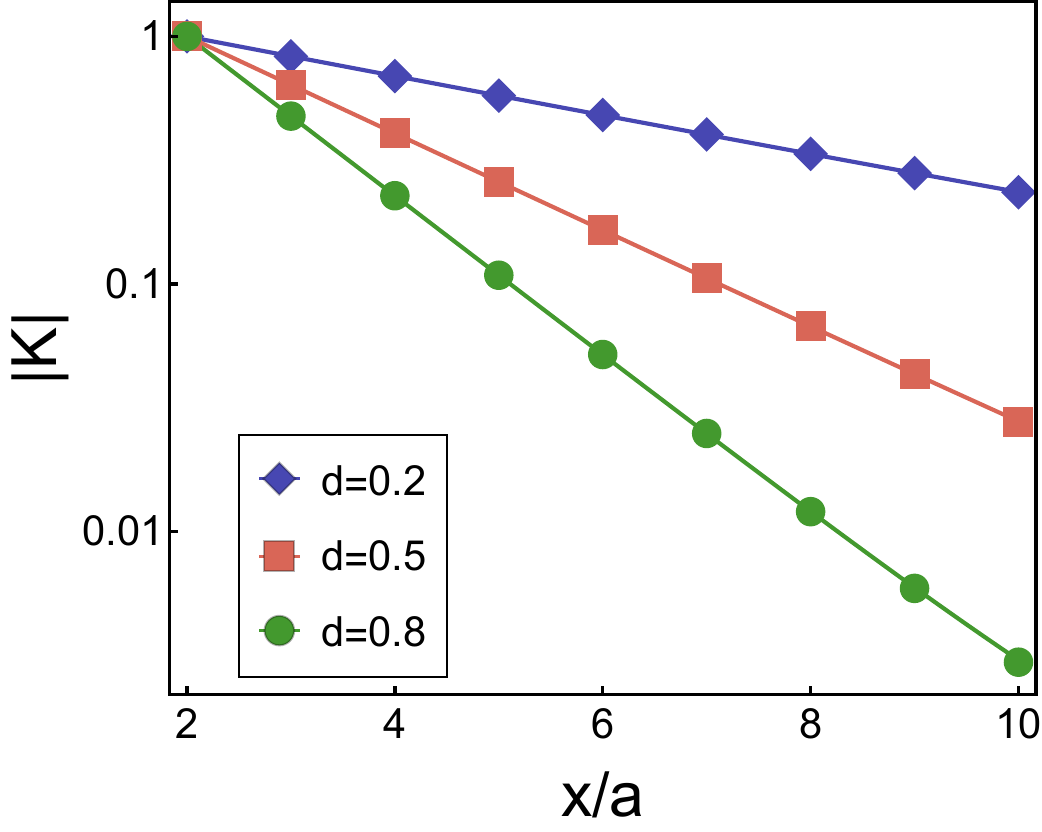} 
\put(0,75){\text{(a)}} 
\end{overpic}
\begin{overpic}[width=0.45\linewidth]{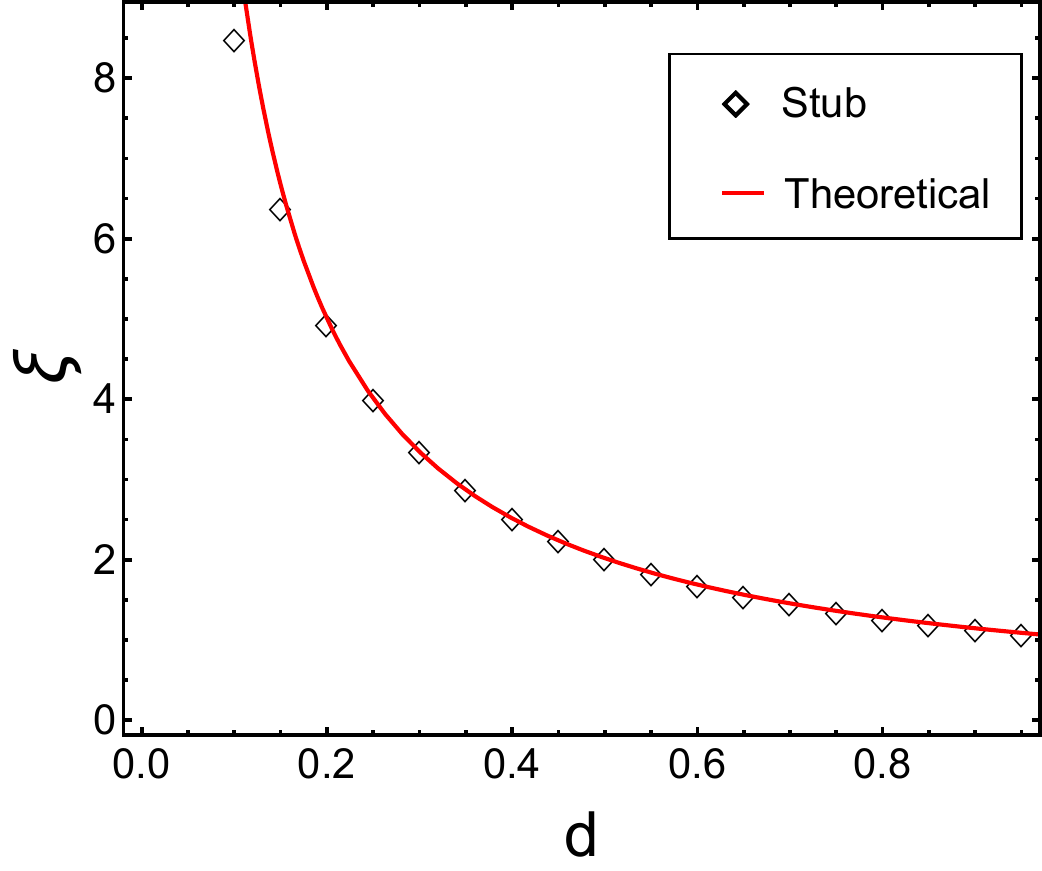} 
\put(-2,78){\text{(b)}} 
\end{overpic}
\caption{\label{fig:stub}(a) Anomalous correlation function $|K_A(x)|$ on the $A$ sublattice in the superconducting Stub lattice for $J=10$ and $U=0.1$. (b) $\ecoh$ (markers) and $\ef$ (red line) as a function of $d$.}
\end{figure}

Using the Bloch basis and keeping only intraband pairing,
\begin{align}\label{eq cf3}
K_{\alpha}(r)
= \frac{1}{N}\sum_{k,n} e^{ikr}\,u_{n\alpha}(k)u_{n\alpha}(-k)\,
\langle c_{kn\uparrow}c_{-kn\downarrow}\rangle.
\end{align}
We employ a self-consistent Bogoliubov–de Gennes (BdG) decoupling,
\begin{align}\label{eq int}
n_{i\alpha,\uparrow}n_{i\alpha,\downarrow}
\simeq \langle n_{i\alpha,\uparrow}\rangle n_{i\alpha,\downarrow}
+ n_{i\alpha,\uparrow}\langle n_{i\alpha,\downarrow}\rangle\\ \nonumber
+ \frac{\Delta_{i\alpha}}{U}c^{\dagger}_{i\alpha}c^{\dagger}_{i\alpha}
+ \frac{\Delta^*_{i\alpha}}{U}c_{i\alpha}c_{i\alpha},
\end{align}
with $\Delta_{i\alpha}=-U\langle c_{i\alpha}c_{i\alpha}\rangle$. Assuming a $k$-independent gap $\Delta$, the expectation value of the pairing amplitude becomes
\begin{align}\label{eq matsu}
\langle c_{kn\uparrow}c_{-kn\downarrow}\rangle
=\frac{1}{\beta}\sum_{\omega_m}\frac{\Delta}{\omega_m^2+(\epsilon_n(k)-\mu)^2+\Delta^2},
\end{align}
where $\omega_m$ is a Matsubara frequency. (The case of $k$-dependent $\Delta$ is discussed in SM.) In the weak-coupling limit with a large normal-state gap separating the flat band from others, the flat band ($\epsilon_{n_{\mathrm{flat}}}(k)-\mu=0$) makes a dominant contribution in Eq.~\eqref{eq matsu} and
\begin{align}\label{eq cf4}
\langle c_{kn\uparrow}c_{-kn\downarrow}\rangle\big|_{n=n_{\mathrm{flat}}}
=\frac{1}{\beta}\sum_{\omega_m}\frac{\Delta}{\omega_m^2+\Delta^2}
=\frac{1}{2}\tanh\!\Bigl(\frac{\beta\Delta}{2}\Bigr).
\end{align}
Hence
\begin{align}\label{eq cff}
K_{\alpha}(r)
\approx \frac{\tanh(\beta\Delta/2)}{2N}\sum_{k} e^{ikr}\,u_{\alpha}(k)u_{\alpha}(-k),
\end{align}
which is basically the Fourier transform of $u_\alpha(k)u_\alpha(-k)$. As in the analysis of $\ef$, the large-$|r|$ decay is set by the analytic region in complex $k$, which is determined by the zeros of $N(k)$ in Eq.~\eqref{eq nkm} and the condition in Eq.~\eqref{eq efconm}. Therefore $K_{\alpha}(r)$ decays in the same way as $\ef$, establishing
\begin{align}
\ecoh=\ef.
\end{align}
This equivalence is corroborated by numerical study of the Stub lattice models [more examples are shown in SM]. Solving the self-consistent BdG equations for Hamiltonian in Eq.~\eqref{eq stubsupm} at $T=0$, we find $|K_A(r)|$ decays exponentially [Fig.~\ref{fig:stub}(a)]. The extracted $\ecoh$ matches well with $\ef$ as a function of $d$ [Fig.~\ref{fig:stub} (b)], with a slight downward deviation at small $d\sim 0.1$ where the normal-state band gap becomes comparable to $\Delta$ and the flat-band projection begins to be invalid.

\textit{Relation between $\ef$ and $\eqm$.---}
Although $\eqm$ and $\ef$ have distinct embedding dependence, they can still be related by placing all orbitals at the same intracell position. For calculational convenience, we also take the CLS overlaps to be real and lattice constant $a=1$. Under these assumptions, we obtain
\begin{align}\label{eq ft1}
\eqm^2\;\leq\;f_{\tmx}(\ef)=\frac{\tmx^2}{4\sqrt{\,1-\frac{1}{\cosh^2(\tmx/\ef)}\,}}
\end{align}
where $\tmx$ is the maximum $|t|$ satisfying $\lambda_t\neq0$ and $f_{\tmx}(\ef)$ is a monotonically increasing function of $\ef$. For large $\ef$, this inequality yields the asymptotic estimate
\begin{align}\label{eq ft2}
  \frac{\ef}{a}
  \;\gtrsim\;
  \frac{4}{\tmx}\,\frac{\eqm^2}{a^2},
\end{align}
showing that $\ef$ is, in general, bounded from below by a quantity proportional to $\eqm^2$ with a prefactor set by $\tmx$.

\textit{Discussion and Conlusion.---}
To conclude, we have introduced an embedding independent, intrinsic length scale $\ef$ of a flat band by using the localization length of an in-gap state proximate to the flat band. We further proved that $\ef$ is fixed solely by CLS overlaps and, in the weak-coupling flat-band superconductor, coincides with the coherence length $\ecoh$. Moreover, $\ef$ admits a lower bound in terms of the QML $\eqm$. Although our explicit construction was carried out in one dimension, the same logic extends to higher dimensions by projecting the lattice along a chosen direction and reducing the problem to an effective one-dimensional one.

Our theory can be extended to the case of nearly flat band $u_\text{n.f.}(k)$ as long as \(w_{\mathrm{n.f.}}\ll\Delta\ll W_{\mathrm{gap}}\) is satisfied where $w_{\mathrm{n.f.}}$ is the bandwidth of $u_\text{n.f.}(k)$ and $W_{\mathrm{gap}}$ indicates the band gap between $u_\text{n.f.}(k)$ and its neighboring band. When the \(k\)-dependence of the pairing amplitude is parametrically weak, and the anomalous correlator is governed solely by the analytic region of \(u_{\mathrm{n.f.,\alpha}}(k)u_{\mathrm{n.f.,\alpha}}(-k)\), we obtain $\xi_{\mathrm{coh}}\simeq\xi_{\mathrm{eff}}$ as before.
Also, the bound between QML and \(\ef\) carries over with small corrections, $\eqm^{2}\;\lesssim\; f_{T}\!\big(\xi_{\mathrm{eff}}\big)$.
[See SM for more detailed discussion.]

Beyond superconductivity, the idea of \(\ef\) can be further applied to various problems related to the flat-band localization, including flat-band magnetism~\cite{PhysRevB.108.235127,Multer2023Imaging} and vortex-bound states in flat-band superconductors~\cite{li2025vortexstatescoherencelengths}, which we leave for future studies.

\begin{acknowledgments}
S. L., S.-H. L., B.-J.Y were supported by Samsung Science and Technology Foundation under project no.  SSTF-BA2002-06, National Research Foundation of Korea (NRF) funded by the Korean government(MSIT),  grant no. RS-2021-NR060087 and RS-2025-00562579,  Global Research Development-Center (GRDC) Cooperative Hub Program through the NRF funded by the MSIT,  grant no. RS-2023-00258359, Global-LAMP program of  the NRF funded by the Ministry of Education, grant no.  RS-2023-00301976.
\end{acknowledgments}

\nocite{*}

\clearpage
\onecolumngrid           
\begin{center}
    \large\textbf{Supplementary Material for "Embedding independent length scale of flat bands"}
\end{center}
\addcontentsline{toc}{section}{Supplemental  material}
\setcounter{section}{0}

\setcounter{page}{1}
\setcounter{equation}{0}
\setcounter{figure}{0}
\setcounter{table}{0}
\renewcommand{\thepage}{S\arabic{page}}
\renewcommand{\theequation}{S\arabic{equation}}
\renewcommand{\thefigure}{S\arabic{figure}}
\renewcommand{\thetable}{S\arabic{table}}

\makeatletter
\setcounter{NAT@ctr}{0}
\makeatother

\section{Embedding dependence of quantum metric length}

Let $|k,\alpha\rangle_{N}=\frac{1}{N_{\rm cell}}\sum_{R}e^{ik(R+\tau_\alpha)}|R,\alpha\rangle$ be the non-periodic Bloch basis and $|k,\alpha\rangle_{P}=\frac{1}{N_{\rm cell}}\sum_{R}e^{ikR}|R,\alpha\rangle$ the periodic one. Denote by $u_{n,\eta}(k)$ ($\eta=N,P$) the corresponding normalized eigenvectors, so that $|\psi_n(k)\rangle=\sum_\alpha u_{n,\eta}(k)_\alpha\,|k,\alpha\rangle_\eta$.

The two conventions are related by a $k$-dependent diagonal unitary:
\begin{align}\label{eq utau}
u_{n,N}(k)_{\alpha}=e^{-ik\tau_\alpha}\,u_{n,P}(k)_{\alpha},
\end{align}
so $u_{n,N}$ inherits explicit dependence on orbital positions $\{\tau_\alpha\}$, while $u_{n,P}$ does not.

Consider a small embedding shift $\tau_\alpha\mapsto\tau_\alpha+\delta\tau_\alpha$. To linear order,
\begin{align}\label{eq upk}
\big|\partial_k u'_{n,N}(k)\big\rangle
=\sum_{\alpha}e^{-ik\delta\tau_\alpha}\big(-i\,\delta\tau_\alpha+\partial_k\big)u_{n,N}(k)_{\alpha}\,|k,\alpha\rangle.
\end{align}
Hence
\begin{align}\label{eq dg1}
\langle \partial_k u'_{n,N}|\partial_k u'_{n,N}\rangle
=\langle \partial_k u_{n,N}|\partial_k u_{n,N}\rangle
+\sum_{\alpha} i\,\delta\tau_\alpha\,u^*_{n,N,\alpha}\,\partial_k u_{n,N,\alpha}+\text{c.c.}+O(\delta\tau^2),
\end{align}
and
\begin{align}\label{eq dg2}
\langle u'_{n,N}|\partial_k u'_{n,N}\rangle
=\langle u_{n,N}|\partial_k u_{n,N}\rangle
-\sum_{\alpha} i\,\delta\tau_\alpha\,|u_{n,N,\alpha}|^2+O(\delta\tau^2).
\end{align}
The quantum metric $g_n(k)=\langle\partial_k u|\,(1-|u\rangle\langle u|)\,\partial_k u\rangle$ then changes by
\begin{align}\label{eq dg3}
\delta g_n(k)
=\sum_{\alpha} i\,\delta\tau_\alpha\Big(\partial_k u_{n,N,\alpha}
-|u_{n,N,\alpha}|^2\,\langle u_{n,N}|\partial_k u_{n,N}\rangle\Big)+\text{c.c.}+O(\delta\tau^2),
\end{align}
which makes explicit the embedding dependence of $g_n(k)$ (and thus of $\eqm=\sqrt{\int \frac{dk}{2\pi}g_n(k)}$) when $u_{n,N}$ is used.

One might wonder whether the integral over $k$ cancels the embedding dependence, as happens for Berry curvature and Chern number. It does not. A simple counterexample is
\begin{align}\label{eq counter1}
u_{1}(k)=\begin{bmatrix} a\,e^{ik/2} \\ b \end{bmatrix},\qquad
u_{2}(k)=\begin{bmatrix} a \\ b \end{bmatrix},
\end{align}
with $|a|^2+|b|^2=1$. The second vector is $k$-independent, so $g_2(k)=0$. For the first, a short calculation gives
\[
g_1(k)=\frac{|a|^2|b|^2}{4},
\]
which is constant in $k$ but nonzero for generic $(a,b)$. Since both metrics are $k$-independent, their integrals (the squared QMLs) differ. Therefore the QML is embedding dependent when computed from the non-periodic Bloch eigenvector.

\section{Localization length of in-gap state}

\subsection{Localization length near a dispersive band}

We first show that the localization length $\xi$ of an in-gap state diverges when its energy approaches a dispersive bulk band, in the perspective of band structure. Let $|\psi_n(k)\rangle$ be an eigenstate of band $n$ with Bloch momentum $k$; it can be expanded as
\begin{align}\label{eq blochvec}
    |\psi_n(k)\rangle
    = \sum_{R,\alpha}
      e^{ikR}\,
      u_{\alpha}(k)\,|R,\alpha\rangle ,
\end{align}
where $R$ labels unit cells, $\alpha$ denotes orbitals, and $u(k)$ is the periodic Bloch vector of $H(k)$ as defined in the main text. To capture exponentially localized in-gap states, we analytically continue $k\to \bk\in\mathbb{C}$. Although $H(\bk)$ is then non-Hermitian, right eigenvectors and eigenvalues satisfy
\begin{align}\label{eq eig}
    H(\bk)\,u(\bk) = E(\bk)\,u(\bk) .
\end{align}
If $E(\bk)\in\mathbb{R}$, the spatial profile of corresponding wavefunction picks up $e^{i\bk x}$, so $\mathrm{Im}\,\bk$ controls exponential decay. The collection of $\bk$ with $E(\bk)\in\mathbb{R}$ forms real-energy curves in the complex-$k$ plane (the \emph{complex-momentum band structure}) and provides candidate decay rates for in-gap solutions~\cite{cpmomentum}.

A physical in-gap state is a linear combination of the eigenvectors $u(\bk_{\mathrm{sol}})$ at all solutions $\{\bk_{\mathrm{sol}}\}$ of $E(\bk_{\mathrm{sol}})=E_{\text{in-gap}}$. Each component decays as $e^{-|\mathrm{Im}\,\bk_{\mathrm{sol}}|x}$, and the slowest-decaying one dominates as $x\to\infty$. Thus the localization length of $\psi$ is
\begin{align}\label{eq xiloc}
    \xi(\psi)
    = \max_{\bk_{\mathrm{sol}}}
      \frac{1}{\bigl|\mathrm{Im}\,\bk_{\mathrm{sol}}\bigr|},
    \qquad
    H(\bk_{\mathrm{sol}})\,u(\bk_{\mathrm{sol}})
    = E_{\text{in-gap}}\,u(\bk_{\mathrm{sol}}),
\end{align}
provided no fine-tuned cancellation removes the leading exponential.

Let the dispersive band have an isolated non-degenerate extremum at $k_0$ with energy $E_0$. Eigenvalues of a matrix $H(\bk)$ remain analytic wherever they are non-degenerate~\cite{analeig}. Analyticity then implies
\[
  E(k)\simeq E_0 \pm \frac{\hbar^2}{2m}(k-k_0)^2
\]
in a neighborhood of $k_0$. The same holds for complex $\bk$. Setting $\bk=k_0+i\kappa$ with $\kappa\in\mathbb{R}$ gives
\[
  E(\bk)=E_0 \mp \frac{\hbar^2}{2m}\kappa^2\in\mathbb{R},
\]
so for $E_{\text{in-gap}}$ close to $E_0$ there are solutions with $|\kappa|\sim \sqrt{2m|E_{\text{in-gap}}-E_0|}/\hbar$. Hence, the imaginary part of the corresponding complex momentum vanishes as the in-gap energy approaches the dispersive band. Thus,
\begin{align}\label{eq disp}
    \lim_{\Delta E\to 0}\xi(\psi)=\infty,
    \qquad
    \Delta E=\bigl|E_{\text{in-gap}}-E_0\bigr|.
\end{align}

\begin{figure}
\centering
\begin{overpic}[width=0.3\linewidth]{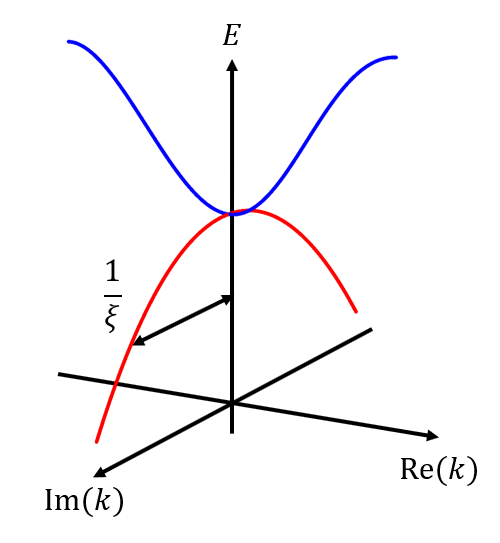}
\put(-2,75){\text{(a)}}
\end{overpic}
\begin{overpic}[width=0.3\linewidth]{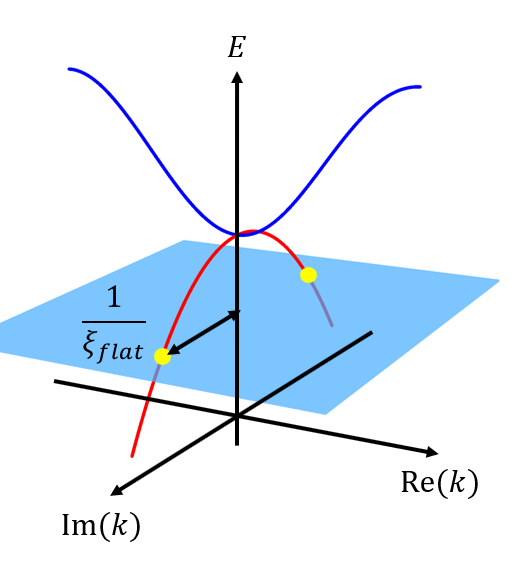}
\put(-2,75){\text{(b)}}
\end{overpic}
\caption{\label{fig:bds}
(a) Isolated dispersive band (blue) and real-energy arc of in-gap solutions in the complex-$k$ plane (red). As the in-gap energy approaches the band minimum, $\mathrm{Im}\,\bk\to 0$ and $\xi\to\infty$.
(b) Flat-band plane (cyan) and crossings (yellow) with real-energy arcs from other bands. Near a flat band, the controlling decay is set by the smallest $|\mathrm{Im}\,\bk|$ among such crossings, leading to a finite $\xi$.
}
\end{figure}

To see that real-energy arcs necessarily emanate from the extremum, expand
\begin{align}\label{eq eapp}
    E(\bk)=E_0+\sum_{i=2}^{\infty}c_i\,(\bk-k_0)^i ,
\end{align}
with real coefficients $c_i$. For $\bk=k_0+\epsilon e^{i\phi}$ with sufficiently small real $\epsilon>0$, the leading contribution to the imaginary part behaves as
$\mathrm{Im}\,E(\bk)\propto \sin(2\phi)$, so it changes sign at least four times as
$\phi$ winds by $2\pi$. This guarantees at least two angles with
$\mathrm{Im}\,E(\bk)=0$ away from the real axis (in addition to two on it), yielding
the desired complex solutions with small $|\mathrm{Im}\,\bk|$.

\subsection{Localization length near a flat band}

In contrast, $\xi$ remains finite when the in-gap energy approaches an isolated flat band. Every flat band admits compact localized states (CLSs) supported on finitely many unit cells. Translation symmetry generates a family $\{|\mathrm{CLS};R\rangle\}$, and their Bloch superposition reads
\begin{align}\label{eq clseig2}
    |\psi_{\text{flat}}(\bk)\rangle
    = \sum_{R} e^{i\bk R}\,|\mathrm{CLS};R\rangle ,
\end{align}
which continues to hold for complex~$\bk$. Thus the flat band spans a real-energy \emph{plane} $E(\bk)\equiv E_{\text{flat}}$ in the complex-$k$ domain [Fig.~\ref{fig:bds}(b)]. Unlike in dispersive bands, this plane does not generate a real-energy arc extending into the gap because all eigenvalues are pinned to the flat band energy, and thus the divergence mechanism is absent. Therefore, regardless of the origin of the in-gap state, the localization length does not diverge as its energy approaches the flat band.

When the in-gap energy approaches $E_{\text{flat}}$, the controlling complex solutions come from other (generically dispersive) bands whose real-energy arcs intersect the flat plane. Among all such intersections, the one with smallest $|\mathrm{Im}\,\bk|$ sets the asymptotic decay via Eq.~\eqref{eq xiloc}, so the in-gap state stays exponentially localized even at $E_{\text{in-gap}}=E_{\text{flat}}$.

Additionally, at least one of the intersections between the flat plane and the real energy curve necessarily has the localization length $\ef$. If a state with flat band energy that is orthogonal to all CLSs is identified, it represents the crossing point between the real energy line of another band and the flat band plane. Let us consider the following state,
\begin{align}
    |\psi'(\bk)\rangle=\sum_n e^{i\bk n}|\text{CLS};n\rangle\, ,
\end{align}
which is the form of Eq.~\eqref{eq clseig2} with the lattice constant set to unity. The inner product between $|\psi'(\bk)\rangle$ and the $m^\text{th}$ CLS is expressed using the overlap function,
\begin{align}\label{eq clsoverlap}
    \lambda_{t}
    = \bigl\langle\mathrm{CLS};n\,\bigl|\,\mathrm{CLS};n-t\bigr\rangle,
    \quad t\in\mathbb{Z}\, ,
\end{align}
as
\begin{align}\label{eq cod}
    \langle \text{CLS};m|\psi'(\bk)\rangle=e^{i\bk m}\sum_t \lambda_t e^{-i\bk t}.
\end{align}
The condition for Eq.~\eqref{eq cod} to be zero is $\sum_t \lambda_t e^{-i\bk t}=0$. This is analogous to the condition that determines $\ef$,
\begin{align}\label{eq efcon}
    \ef=\max\Bigl\{\frac{1}{|\text{Im}(k)|}:\sum_{t\in\mathbb{Z}}\lambda_{t}\,e^{ikt}=0\Bigr\}.
\end{align}
By attaching a minus sign to the $k$ that determines $\ef$, one can find solution of $\sum_{t} \lambda_{t}\, e^{-i\mathbf{k}t} = 0$, with same localization length $\ef$. Under this condition, $|\psi(\bk)\rangle$ is orthogonal to all CLSs, which means that it is not a superposition of CLSs while still possessing the flat band energy. This corresponds to a situation in the complex band structure where the state is degenerate at the flat-band energy. Therefore, there exists a intersection point in flat band plane which possesses the localization length $\ef$.

It should be noted, however, that not every degenerate point on the flat-band plane yields the localization length $\xi_{\text{flat}}$. Other accidental degenerate points may yield arbitrary localization lengths, $1/|\mathrm{Im}\,k|$, and such points can determine the localization length of an in-gap state proximate to the flat band. What we have established is that there always exists at least one solution whose localization length is exactly $\xi_{\text{flat}}$.

\subsection{Decay of in-gap states induced by local perturbation}

We now show that the localization length of an in-gap state induced by a local perturbation is set by the localization length of the bare Green's function evaluated at the same energy.

We start from the bare Green's function of $H_0$,
\begin{align}\label{eq green0}
G^{0}(i\alpha,j\beta;\omega)
&=\frac{1}{N}\sum_{n,k}
\frac{\langle i\alpha |\psi_{n}(k)\rangle
      \langle \psi_{n}(k)|j\beta\rangle}
     {\omega + i\eta - E_{n}(k)} \\ \nonumber
&=\frac{1}{N}\sum_{n,k}
\frac{u_{n,\alpha}(k)\,u^{*}_{n,\beta}(k)}
     {\omega + i\eta - E_{n}(k)}
\,e^{ik(R_{i}-R_{j})}\, .
\end{align}
Here $| \psi_n(k)\rangle$ are the Bloch eigenstates with band energy $E_n(k)$, and $u_{n,\alpha}(k)$ denotes the periodic Bloch eigenfunction with $\alpha$ orbital component.

For a given real energy $\omega$ in the spectral gap of $H_0$, we define the sharp exponential decay rate of the bare Green's function by
\begin{align}
a_\star(\omega)\;:=\;\limsup_{|r|\to\infty}\frac{-\log|G^0(i\alpha,j\beta;\omega)|}{|r|}\,,
\qquad
r = R_i - R_j .
\end{align}
Physically, $a_\star(\omega)$ is the inverse localization length of the bare propagator at energy $\omega$, i.e. $|G^0(i\alpha,j\beta;\omega)| \sim e^{-a_\star(\omega)|R_i-R_j|}$ at large separation.

Now consider a local perturbation $V$ that is nonzero only on a finite number of unit cells near $R=0$, and define $H = H_0 + V$. Suppose $H$ hosts an in-gap eigenstate $|\psi\rangle$ with energy $\omega_b$:
\begin{align}
H |\psi\rangle = \omega_b |\psi\rangle,\qquad \omega_b \text{ lies in a band gap of } H_0.
\end{align}
From the eigenvalue equation we obtain the standard Lippmann--Schwinger form
\begin{align}
|\psi\rangle
= G^0(\omega_b)\,V|\psi\rangle,
\qquad\Rightarrow\qquad
\psi_\alpha(R_i)
=\sum_{j\beta}
G^0(i\alpha,j\beta;\omega_b)\,
\langle j\beta|V|\psi\rangle.
\end{align}
This sum is finite, because $V$ acts only on a finite set $S$ of sites (or orbitals) near the origin.

By the definition of $a_\star(\omega_b)$, for any $\varepsilon>0$ there exists $C_\varepsilon$ such that, for $|R_i-R_j|$ sufficiently large,
\begin{align}
|G^0(i\alpha,j\beta;\omega_b)|
\;\le\;
C_\varepsilon \, e^{-(a_\star(\omega_b)-\varepsilon)\,|R_i-R_j|}.
\end{align}
Using this in the above expression for $\psi_\alpha(R_i)$ gives
\begin{align}
|\psi_\alpha(R_i)|
&\le
\sum_{j\beta}
|G^0(i\alpha,j\beta;\omega_b)|\,
\big|\langle j\beta|V|\psi\rangle\big| \\
&\le
C_\varepsilon \,
e^{-(a_\star(\omega_b)-\varepsilon)\,|R_i|}
\sum_{j\beta\in S}
e^{(a_\star(\omega_b)-\varepsilon)\,|R_j|}\,
\big|\langle j\beta|V|\psi\rangle\big|.
\end{align}
Since $S$ is finite, the sum over $(j,\beta)$ is a finite constant independent of $R_i$. We can therefore absorb it into a prefactor and write
\begin{align}
|\psi_\alpha(R_i)|
\;\le\;
C'_\varepsilon\,
e^{-(a_\star(\omega_b)-\varepsilon)\,|R_i|}.
\end{align}
Letting $\varepsilon\downarrow 0$ we obtain the asymptotic upper bound
\begin{align}
|\psi_\alpha(R_i)|
\;\le\;
C' \, e^{-a_\star(\omega_b)\,|R_i|}.
\end{align}
This shows that the in-gap eigenstate $|\psi\rangle$ cannot decay more slowly than the bare Green's function at the same energy. Also in generic (non-fine-tuned) situations, the leading exponential factor in $\psi_\alpha(R_i)$ is in fact set exactly by $a_\star(\omega_b)$, since any additional suppression would require exact destructive interference among a finite number of terms in the sum. Thus, up to such nongeneric cancellations, the localization length of the bound state $|\psi\rangle$ coincides with that of the bare Green's function at the same energy.

We now connect this general statement to the standard single-impurity analysis. Consider a single point impurity introduced at lattice site $R_{i_0}$ and orbital $\alpha_0$,
\begin{align}
V \;=\; V_0\,c_{i_0\alpha_0}^{\dagger}c_{i_0\alpha_0}\,.
\end{align}
The full Green’s function satisfies the Dyson equation
\begin{align}
G = G^{0} + G^{0}\,T\,G^{0}, 
\qquad
T(\omega)=\bigl[1 - V\,G^{0}(\omega)\bigr]^{-1}V.
\end{align}
The corresponding $T$-matrix is given by
\begin{align}
T(\omega)
=|i_0 \alpha_0\rangle\frac{V_0}{1 - V_0\,G^{0}(i_0\alpha_0,i_0\alpha_0;\omega)}\langle i_0 \alpha_0|,
\end{align}
and the bound-state energy $\omega_{b}$ is determined by the pole condition
\begin{align}\label{polecon}
1 - V_0\,G^{0}(i_0\alpha_0,i_0\alpha_0;\omega_b) = 0 .
\end{align}

The full Green’s function at energy $\omega$ can then be written as
\begin{align}
G(i\alpha,i_0\alpha_0;\omega)
=G^0(i\alpha,i_0\alpha_0;\omega)
+G^0(i\alpha,i_0\alpha_0;\omega)
\frac{V_0}{1 - V_0\,G^{0}(i_0\alpha_0,i_0\alpha_0;\omega)}
G^0(i_0\alpha_0,i_0\alpha_0;\omega).
\end{align}
When $\omega$ lies within the band gap, the pole of $G(i\alpha,i_0\alpha_0;\omega)$ originates from the term
$\frac{V_0}{1 - V_0\,G^{0}(i_0\alpha_0,i_0\alpha_0;\omega)}$, since $G^0(i\alpha,i_0\alpha_0;\omega)$ itself has no pole.
Expanding the denominator of this term to first order around $\omega = \omega_b$ and using Eq.~\eqref{polecon}, we obtain
\begin{align}\label{polecon2}
\frac{V_0}{1 - V_0\,G^{0}(i_0\alpha_0,i_0\alpha_0;\omega)}
\;\approx\;
\frac{1}{-(\omega-\omega_b)\,G^{0\prime}(i_0\alpha_0,i_0\alpha_0;\omega_b)},
\end{align}
where the prime in $G^{0\prime}(i_0\alpha_0,i_0\alpha_0;\omega_b)$ denotes differentiation with respect to energy $\omega=\omega_b$.
Substituting this into the Dyson equation yields
\begin{align}
G(i\alpha,i_0\alpha_0;\omega)
\;\approx\;
G^0(i\alpha,i_0\alpha_0;\omega)
-\frac{G^0(i\alpha,i_0\alpha_0;\omega_b)\,G^0(i_0\alpha_0,i_0\alpha_0;\omega_b)}
{(\omega-\omega_b)\,G^{0\prime}(i_0\alpha_0,i_0\alpha_0;\omega_b)}.
\end{align}
The residue of $G(i\alpha,i_0\alpha_0;\omega)$ at the pole $\omega=\omega_b$ can thus be identified as the coefficient of
$\frac{1}{\omega-\omega_b}$,
\begin{align}\label{eq gres}
\text{Res}\{G(i\alpha,i_0\alpha_0;\omega_b)\}
=-\frac{G^0(i\alpha,i_0\alpha_0;\omega_b)\,G^0(i_0\alpha_0,i_0\alpha_0;\omega_b)}
{G^{0\prime}(i_0\alpha_0,i_0\alpha_0;\omega_b)}\, .
\end{align}
The residue of the Green’s function at the bound-state pole also corresponds to the outer product of the bound-state wavefunction, as follows from the spectral representation
\begin{align}
G(i\alpha, j\beta; \omega) = \sum_n \frac{\psi_{n,i\alpha}\,\psi^{*}_{n,j\beta}}{\omega - E_n}.
\end{align}
Hence,
\begin{align}\label{eq gres2}
\text{Res}\{G(i\alpha,i_0\alpha_0;\omega_b)\}
=\psi_{i\alpha}(\omega_b)\,\psi^{*}_{i_{0}\alpha_{0}}(\omega_b),
\end{align}
where $\psi_{i\alpha}(\omega_b)=\langle i\alpha |\psi(\omega_b)\rangle$.
Comparing Eq.~\eqref{eq gres} and Eq.~\eqref{eq gres2} with respect to the variable $i\alpha$, we obtain
\begin{align}\label{eq ingapstate}
\psi_{i\alpha}(\omega_b)=-G^0(i\alpha,i_0\alpha_0;\omega_b)\frac{G^0(i_0\alpha_0,i_0\alpha_0;\omega_b)}
{G^{0\prime}(i_0\alpha_0,i_0\alpha_0;\omega_b)\psi^{*}_{i_{0}\alpha_{0}}(\omega_b)}=\frac{C}{N}\sum_{n,k}
\frac{u_{n,\alpha}(k)\,u^{*}_{n,\alpha_{0}}(k)}
     {\omega_{b} - E_{n}(k)}
\,e^{ik(R_{i}-R_{i_{0}})}
\end{align}
where the constant $C=-G^0(i_0\alpha_0,i_0\alpha_0;\omega_b)/
G^{0\prime}(i_0\alpha_0,i_0\alpha_0;\omega_b)\psi^{*}_{i_{0}\alpha_{0}}(\omega_b)$.

\subsection{Paley-Wiener theorem}

The Paley–Wiener theorem~\cite{Paley1934} used to obtain the exponential decaying factor of the in-gap state in the main text is as follows.
\begin{theorem}[Paley-Wiener]
Let $f(z)$ be $2\pi$-periodic. Then $f(z)$ admits a holomorphic extension to the strip
$|\text{Im}(z)|<\gamma$ iff its Fourier coefficients satisfy
\[
|c_n|\;\le\;C_\varepsilon e^{-(\gamma-\varepsilon)|n|},\quad \forall\,\varepsilon>0.
\]
Moreover, the maximal width of the analytic strip is determined by the exponential decay rate:
if $f$ does not extend beyond $|\Im z|<\gamma$, then one has
\[
\limsup_{|n|\to\infty}\frac{1}{|n|}\ln\frac{1}{|c_n|}=\gamma,
\]
so that decay faster than $e^{-\gamma|n|}$ is impossible.
\end{theorem}
By applying the Paley–Wiener theorem to \(f=u_{\alpha}(k)u^{*}_{\alpha_{0}}(k)\), we obtain the result quoted in the main text. Here we explain the analytic continuation in more detail.

For real \(k\), \(u_{\alpha}(k)u^{*}_{\alpha_{0}}(k)\) can be written as
\begin{align}\label{eq fourfuncs}
    u_{\alpha}(k)\,u^{*}_{\alpha_0}(k) 
    =\frac{1}{|N(k)|^2}
      \Bigl(\sum_{R}
      \langle R,\alpha|\mathrm{CLS};0\rangle\,e^{-ikR}\Bigr)
      \Bigl(\sum_{R'}
      \langle \mathrm{CLS};0|R',\alpha_0\rangle\,e^{ikR'}\Bigr),
\end{align}
where \(N(k)^2 = \sum_{t\in\mathbb{Z}}\lambda_t e^{ikt}\). Since \(\lambda_{-t}=\lambda_t^{*}\), the sum \(N(k)^2\) is real for real \(k\), and thus \(|N(k)|^2 = N(k)^2\) holds on the real axis. By replacing \(|N(k)|^2\) by \(N(k)^2\) in Eq.~\eqref{eq fourfuncs}, one can yield analytic function
\begin{align}\label{eq fourfuncs2}
    \frac{1}{N(k)^2}
      \Bigl(\sum_{R}
      \langle R,\alpha|\mathrm{CLS};0\rangle\,e^{-ikR}\Bigr)
      \Bigl(\sum_{R'}
      \langle \mathrm{CLS};0|R',\alpha_0\rangle\,e^{ikR'}\Bigr).
\end{align}
This function therefore defines the analytic continuation of \(u_{\alpha}(k)u^{*}_{\alpha_0}(k)\), whose analytic region is bounded by the zeros of \(N(k)^2\), i.e., by the solutions of \(\sum_{t\in\mathbb{Z}}\lambda_t e^{ikt}=0\).

\subsection{$\ef$ of several 1D flat-band systems}

\begin{figure}
\centering
\begin{overpic}[width=0.3\linewidth]{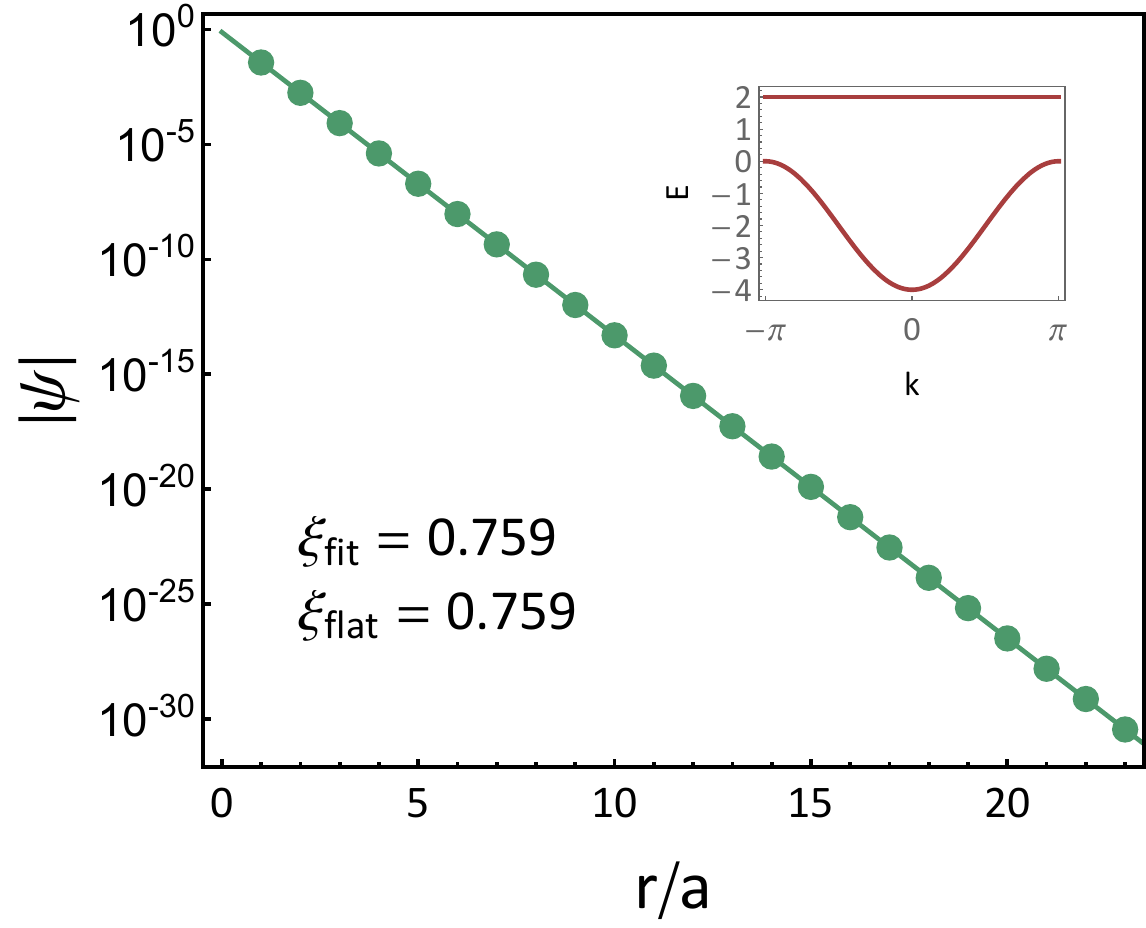}
\put(-2,75){\text{(a)}}
\end{overpic}
\begin{overpic}[width=0.3\linewidth]{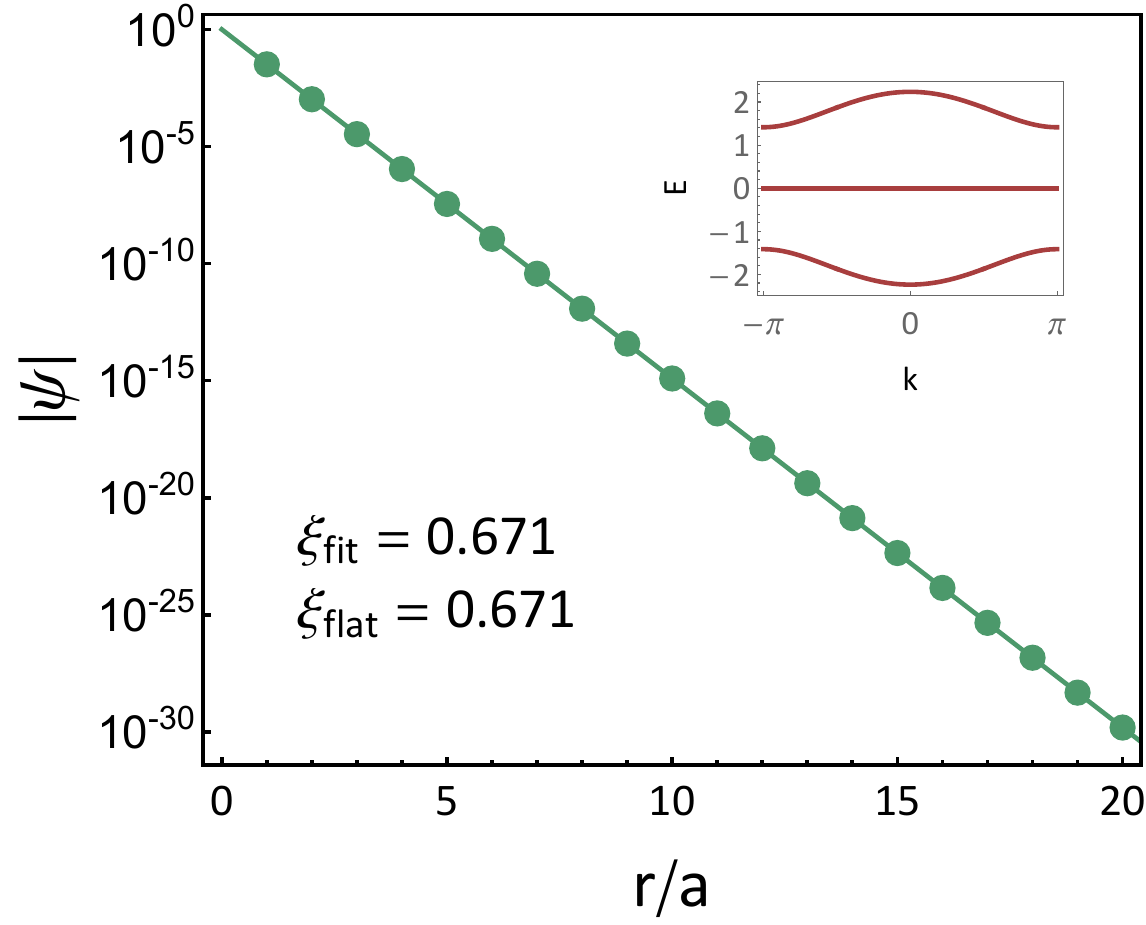}
\put(-2,75){\text{(b)}}
\end{overpic}
\begin{overpic}[width=0.3\linewidth]{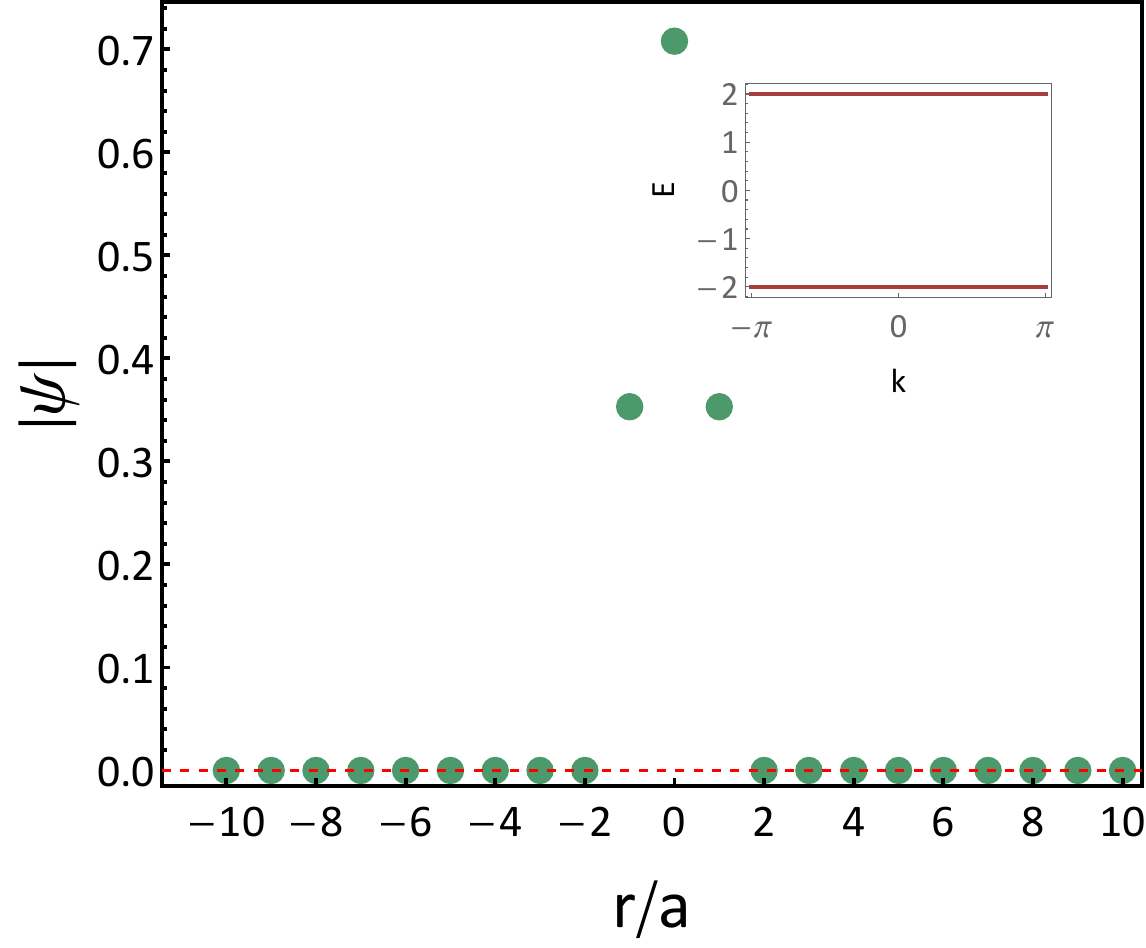}
\put(-2,75){\text{(b)}}
\end{overpic}
\caption{\label{fig:num}(a) Numerical result for the sawtooth lattice. The localization length of an in-gap state near the flat band is extracted by fitting. (b) Numerical result for the 1D Lieb lattice. (c) Numerical result for the Creutz ladder.
}
\end{figure}

For representative one-dimensional flat-band systems, i.e., the sawtooth lattice, the 1D Lieb lattice, and the Creutz ladder, we have calculated $\xi_{\text{flat}}$ and verified the results numerically.

The periodic $k$-space Hamiltonian of the sawtooth lattice is
\begin{align}\label{eq swath}
H(k)=\begin{bmatrix}
-2t\cos(k)&-\sqrt{2}t(1+e^{-ik}) \\
-\sqrt{2}t(1+e^{ik})&0
\end{bmatrix},
\end{align}
with the lattice constant set to unity.
This model has two orbitals, $\{A,B\}$, and one flat band. The normalized CLS can be written as
\begin{align}\label{eq swatcls}
|\text{CLS};R\rangle=\frac{1}{\sqrt{2}}|A,R\rangle-\frac{1}{2}|B,R-1\rangle-\frac{1}{2}|B,R\rangle,
\end{align}
and the periodic Bloch eigenvector of the flat band is
\begin{align}\label{eq swatev}
u(k)=\frac{1}{N(k)}\begin{bmatrix}
\frac{1}{\sqrt{2}}\\[4pt]
-\tfrac{1}{2}(1+e^{ik})
\end{bmatrix},
\end{align}
with the normalization factor $N(k)$. In this case, only the nearest-neighbor CLS overlaps are nonzero, $\lambda_1=\lambda_{-1}=\tfrac{1}{4}$. From the expression given in the main text,
\begin{align}\label{eq loclen2}
    \xi_{\text{flat}}
    \;=\;
    \Bigl[\,
      \operatorname{arccosh}\!\Bigl(\tfrac{1}{|2\lambda_{1}|}\Bigr)
    \Bigr]^{-1},
\end{align}
we obtain $\xi_{\text{flat}}=\tfrac{1}{\operatorname{arccosh}(2)}$. This analytic result is confirmed numerically [Fig.~\ref{fig:num}(a)], from the in-gap state at $E=2.00423$, which appears when an onsite potential $V_0=0.01$ is applied to a single $B$ orbital in the Hamiltonian with hopping amplitude $t=1$.

The Hamiltonian of the 1D Lieb lattice is
\begin{align}\label{eq lieb}
H(k)=\begin{bmatrix}
0 & (1+d) + (1-d)e^{-ik} & 0 \\
(1+d) + (1-d)e^{ik} & 0 & 2d \\
0 & 2d & 0
\end{bmatrix}.
\end{align}
This lattice has three orbitals, $\{A,B,C\}$, and one flat band. The normalized CLS is
\begin{align}\label{eq liebcls}
|\text{CLS};R\rangle=\frac{1}{\sqrt{2+6d^2}}\Big\{2d|A,R\rangle-(1-d)|C,R-1\rangle-(1+d)|C,R\rangle\Big\},
\end{align}
and the periodic Bloch eigenvector of the flat band is
\begin{align}\label{eq liebev}
u(k)=\frac{1}{N(k)\sqrt{2+6d^2}}\begin{bmatrix}
2d\\[4pt]
-(1-d)e^{ik}-(1+d)
\end{bmatrix},
\end{align}
with normalization factor $N(k)$. Here, only the nearest-neighbor CLS overlaps are nonzero, $\lambda_1=\lambda_{-1}=\tfrac{1-d^2}{2+6d^2}$. The corresponding localization length is
\[
\xi_{\text{flat}}
=\Biggl[\operatorname{arccosh}\!\Bigl(\tfrac{1+3d^2}{1-d^2}\Bigr)\Biggr]^{-1}.
\]
This analytic result is also confirmed numerically [Fig.~\ref{fig:num}(b)]. For $d=0.5$, the fitted in-gap state with $E=0.00684$ arises from applying an onsite potential $V_0=0.01$ at a $C$ orbital.

Finally, we consider the Creutz ladder, whose Hamiltonian is
\begin{align}\label{eq crl}
H(k)=\begin{bmatrix}
-2t\sin(k) & -2t\cos(k) \\
-2t\cos(k)&2t\sin(k)
\end{bmatrix},
\end{align}
which hosts two nondegenerate flat bands. The corresponding CLS takes the form
\begin{align}\label{eq crlcls}
|\text{CLS};R\rangle=\frac{1}{\sqrt{2}}|A,R\rangle+\frac{i}{\sqrt{2}}|B,R\rangle,
\end{align}
and is strictly confined within a single unit cell. Thus, only $\lambda_{0}$ is nonzero, while all other CLS overlaps vanish. Consequently, there is no solution to the condition
\begin{align}\label{eq efcon}
    \xi_{\text{flat}}=\max\Bigl\{\tfrac{1}{|\text{Im}(k)|}:\sum_{t\in\mathbb{Z}}\lambda_{t}\,e^{ikt}=0\Bigr\}.
\end{align}
However, if this situation is regarded as the limit of Eq.~\eqref{eq efcon} where $\lambda_{t}$ with $t\neq 0$ vanish, then $\xi_{\text{flat}}$ converges to zero. From a physical point of view, if there is no overlap between CLSs, the length scale $\xi_{\text{flat}}$, which measures propagation through overlaps, must vanish. Indeed, a numerical investigation of the in-gap state confirms that it exhibits zero localization length, as shown in Fig.~\ref{fig:num}(c).

\section{Proof of $\xi_{\mathrm{coh}}=\ef$ in flat-band superconductors}

We start from the Hamiltonian with on-site Hubbard interaction,
\begin{align}
H=\sum_{ij\alpha\beta\sigma}
t_{ij,\alpha\beta}\,
c^{\dagger}_{i\alpha\sigma}c_{j\beta\sigma}
-\mu N
+U\sum_{i\alpha}n_{i\alpha\uparrow}n_{i\alpha\downarrow}.
\end{align}
The anomalous correlator is
\begin{align}
K_{\alpha}(r_j-r_i)
=\langle c_{i\alpha\uparrow}c_{j\alpha\downarrow}\rangle
\sim e^{-|r|/\xi_{\mathrm{coh}}}
\quad (|r|\to\infty),
\end{align}
and can be written in momentum space as
\begin{align}
\langle c_{i\alpha\uparrow}c_{j\alpha\downarrow}\rangle
=\frac{1}{N_\text{cell}}\sum_{knm}
e^{ik(R_i-R_j)}u_{n\alpha}(k)u_{m\alpha}(-k)
\langle c_{kn\uparrow}c_{-km\downarrow}\rangle .
\end{align}
Since we assume that superconducting gap satisfies $\Delta\ll W_\text{gap}$, where $W_\text{gap}$ is the energy separation from the dispersive bands, the anomalous interband terms are suppressed by the large energy denominator, and the pairing is therefore dominated by the intraband component of the flat band. Using the approximation
$\langle c_{kn\uparrow}c_{-km\downarrow}\rangle
\simeq\delta_{nm}\langle c_{kn\uparrow}c_{-kn\downarrow}\rangle$,
we obtain
\begin{align}\label{eq cf3}
K_{\alpha}(r) = \frac{1}{N_\text{cell}} \sum_{kn} e^{ikr}u_{n\alpha} (k)u_{n\alpha} (-k)\langle c_{kn\uparrow}c_{-kn\downarrow}\rangle.
\end{align}
The correlation function of Eq.~\eqref{eq cf3} can be written using Matsubara sum as
\begin{align}
\langle c_{kn\uparrow}c_{-kn\downarrow}\rangle=\frac{1}{\beta}\sum_{m}\frac{\Delta_n(k)}{\omega_m^2+(\epsilon_n(k)-\mu)^2+\Delta_n(k)^2}
\end{align}
where $\Delta_n(k)=\sum_{\alpha}\Delta_{\alpha}\,u_{n\alpha}(k)\,u_{n\alpha}(-k)$. Since we set $\mu$ to the flat band energy, the correlation $\langle c_{kn\uparrow}c_{-kn\downarrow}\rangle$ of the flat band
\begin{align}\label{eq cf4}
\langle c_{kn\uparrow}c_{-kn\downarrow}\rangle|_{n=n_{\text{flat}}}=\frac{1}{\beta}\sum_{m}\frac{\Delta(k)}{\omega_m^2+\Delta(k)^2}=\frac{1}{2}\tanh(\frac{\beta\Delta(k)}{2})
\end{align}
is more dominant than that of other bands. Assuming that the band gap is large enough to consider only the flat band, $K_\alpha(r)$ can be approximated as
\begin{align}\label{eq:acf-final}
K_{\alpha}(r)
\simeq 
\frac{1}{2N_{\mathrm{cell}}}
\sum_k e^{ikr}\,
u_{\alpha}(k)u_{\alpha}(-k)\,
\tanh\!\left[\frac{\beta\Delta(k)}{2}\right].
\end{align}
where $u(k)$ is periodic Bloch vector of the flat band and $\Delta(k)$ is $\Delta_{n_\text{flat}}(k)$.

The integrand in Eq.~\eqref{eq:acf-final} factorizes, so $K_\alpha(r)$ is the convolution of the Fourier transforms
\[
\widetilde f_u(r)\equiv\mathcal{F}[u_{\alpha}(k)u_{\alpha}(-k)],
\qquad
\widetilde F(r)\equiv \mathcal{F}[\tanh(\beta\Delta(k)/2)].
\]

Let $\widetilde f_u(r)\sim e^{-|r|/\xi_u}$ and $\widetilde F(r)\sim e^{-|r|/\xi_F}$.
Since the convolution of two exponentially decaying functions is itself exponentially decaying with the slower rate, the asymptotic behavior of $K_\alpha(r)$ is governed by the larger of the two decay lengths, $\xi_u$ and $\xi_F$:
\[
K_\alpha(r)\sim e^{-|r|/\xi_{\mathrm{coh}}},\qquad
\xi_{\mathrm{coh}}=\max\{\xi_u,\xi_F\}.
\]

\subsection{Decay length $\xi_u$ and $\xi_F$}

For large systems, the sum becomes a Brillouin-zone integral,
\[
\widetilde f_u(r)
=\int_{-\pi}^{\pi}\frac{dk}{2\pi}
\,e^{ikr}u_{\alpha}(k)u_{\alpha}(-k).
\]
By the Paley--Wiener theorem, its decay is controlled by the width of the analytic region in complex~$k$ where $u_{\alpha}(k)u_{\alpha}(-k)$ remains analytic.

Using the representation of a flat-band Bloch eigenstate in terms of CLS,
\[
u_{\alpha}(k)
=\frac{e^{i\phi(k)}}{N(k)}
\sum_{R}
\langle R,\alpha|\mathrm{CLS};0\rangle
\,e^{-ikR},
\]
the analyticity of \(u_{\alpha}(k)u_{\alpha}(-k)\) is governed by \(\bigl[N(k)N(-k)\bigr]^{-1}\), where \(N(k) = \bigl[\sum_{t\in\mathbb{Z}}\lambda_{t}\,e^{ikt}\bigr]^{1/2}\).
The zeros of \(N(k)^2\) and \(N(-k)^2\) are related by \(k\to -k\) and therefore have the same distance to the real axis, so \(N(k)\) and \(N(-k)\) share the same analytic-region width.
In any region where \(N(k)\neq 0\) and \(N(-k)\neq 0\), a holomorphic branch of \(N(k)\) and \(N(-k)\) can be chosen, and the choice of branch does not affect analyticity. Consequently, the analytic region of \(u_{\alpha}(k)u_{\alpha}(-k)\) is the same as that of \(1/N(k)^{2}\), and we obtain \(\ef=\xi_u\).

Now consider $\xi_F$.  
Since
\[
F(k)=\tanh\!\left[\frac{\beta\Delta(k)}{2}\right],
\]
its singularities arise from
(i)~non-analyticities of $\Delta(k)$ and
(ii)~poles of $\tanh z$ at  
$z=i\pi(2\ell+1)/2$. From the relation $\Delta(k)=\sum_{\alpha}\Delta_{\alpha}u_{\alpha}(k)u_{\alpha}(-k)$, non-analytic point of $\Delta_n(k)$ is equal to that of $u_{\alpha}(k)u_{\alpha}(-k)$.

Thus $F(k)$ is analytic in the same strip as $\Delta(k)$ unless the equation
\[
\Delta(k)=i\pi k_BT(2\ell+1)
\]
has a solution inside the region. However, it is non-generic for $\Delta(k)$ to take the specific complex values $i\pi k_B T(2\ell+1)$ inside the analytic region, as this requires fine tuning. Each of these conditions imposes a single complex equation and therefore corresponds to a codimension-two constraint in the microscopic parameter space, making such solutions unstable under generic perturbations.
Therefore the singularities of
\[
\mathcal{F}_\alpha(k)=u_{\alpha}(k)u_{\alpha}(-k)F(k)
\]
coincide with those of $u_{\alpha}(k)u_{\alpha}(-k)$, and hence
\[
\xi_F=\xi_u=\ef .
\]

As a result, both contributions to $K_\alpha(r)$ have the same analytic region width, set by the nearest complex zero of $N(k)$ that determines $\ef$.
Thus,
\begin{align}
\xi_{\mathrm{coh}}=\max\{\xi_u,\xi_F\}=\ef.
\end{align}
We therefore conclude that, in a generic uniform fully gapped $s$-wave flat-band superconductor, the coherence length is entirely determined by the analytic structure of the flat-band Bloch eigenstates and equals $\ef$.

\subsection{Numerical calculation of $\ecoh$}

\begin{figure}
\centering
\begin{overpic}[width=0.3\linewidth]{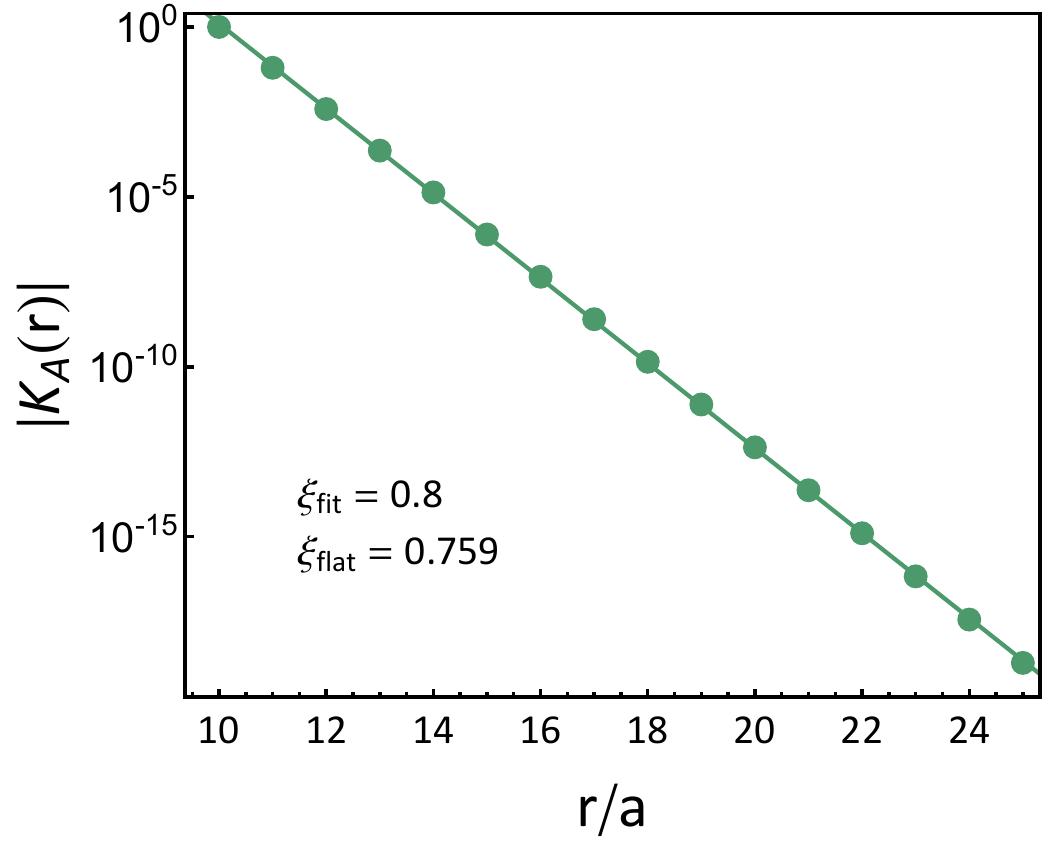}
\put(-2,75){\text{(a)}}
\end{overpic}
\begin{overpic}[width=0.3\linewidth]{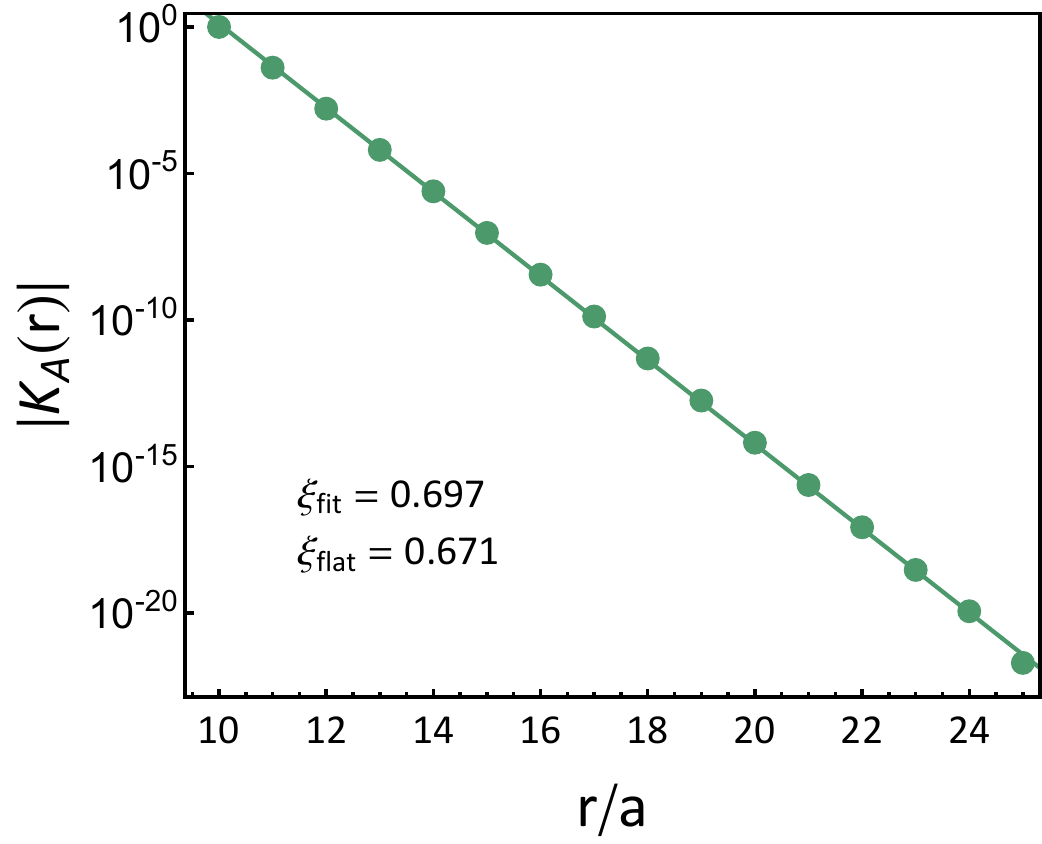}
\put(-2,75){\text{(b)}}
\end{overpic}
\begin{overpic}[width=0.3\linewidth]{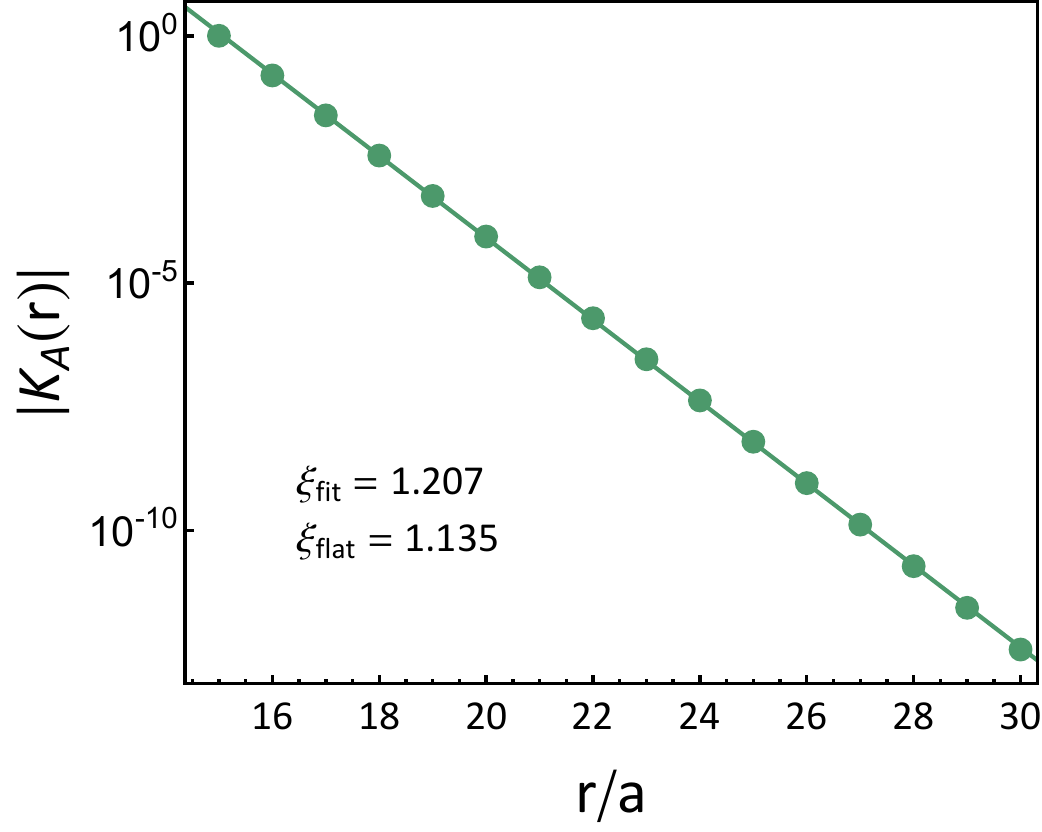}
\put(-2,75){\text{(b)}}
\end{overpic}
\caption{\label{fig:num2}(a) Numerically fitted $|K_A(r)|$ and $\ecoh$ for the sawtooth lattice, (b) for the 1D Lieb lattice, and (c) for the modified diamond lattice.
}
\end{figure}

We perform numerical calculations for three flat-band models: the sawtooth lattice, the one-dimensional Lieb lattice, and a modified diamond lattice. The results are summarized in Fig.~\ref{fig:num2}.

The periodic $k$-space Hamiltonian of the sawtooth lattice is given in Eq.~\eqref{eq swath}, and its CLS and $\ef$ were obtained in the previous section. We set the chemical potential to the flat-band energy, $E=2$, and choose the interaction strength $U=-0.01$. We then compute the anomalous correlator on sublattice $A$ and extract its localization length via an exponential fit, obtaining a value in good agreement with $\ef$. The small deviation can be attributed to the flat-band projection and the neglect of interband pairing.

The Hamiltonian of the 1D Lieb lattice is again given by Eq.~\eqref{eq lieb}. Here, we set the chemical potential to the flat-band energy, $E=0$, and use the same interaction strength, $U=-0.01$. The resulting coherence length $\ecoh$ is found to follow $\ef$ with similar accuracy.

The modified diamond lattice is introduced to demonstrate the validity of our results in a case where the CLS overlaps are complex. Although $\ecoh$ is derived from $N(k)N(-k)$ rather than $|N(k)|^2$, we previously argued that they share the same non-analytic structure, since the non-analytic points appear in sign-reversed pairs. This expectation is confirmed by explicit numerical analysis. The Hamiltonian of the modified diamond lattice is
\begin{align}\label{eq dia}
H(k)=\begin{bmatrix}
0 & 1+ie^{-ik} & 0 \\
1-ie^{ik}& 0 & 1+e^{ik} \\
0 & 1+e^{-ik} & 0
\end{bmatrix}.
\end{align}
It hosts a flat band with CLS
\begin{align}\label{eq diacls}
|\text{CLS};R\rangle=\frac{1}{2}|A,R\rangle+\frac{i}{2}|C,R\rangle+\frac{1}{2}|A,R+1\rangle-\frac{1}{2}|C,R+1\rangle,
\end{align}
for which the nearest-neighbor CLS overlap takes a complex value, $\lambda_{\pm1}=(1\pm i)/4$, and $\ef=1/\text{arccosh}(\sqrt{2})$. With $\mu=0$ and $U=-0.01$, the numerically obtained $\ecoh$ agrees with $\ef$ up to small corrections.

\section{Relation between $\ef$ and $\eqm$}

In this section, we derive the function $f_{\tmx}$ that characterizes the relation between $\ef$ and $\eqm$. Because $\ef$ and $\eqm$ have different embedding dependence in general, we first place all orbitals at the same intracell position to unify that dependence. Under this convention, we ask the following question: for a given CLS overlaps, how large can $\eqm$ possibly be? The answer will give an upper bound on $\eqm$ in terms of $\ef$.

For a given flat band and its set of CLS overlaps $\{\lambda_t\}$, we find
\begin{align}\label{eqmmax}
    \eqm^2 \leq \xi_{\text{QM,max}}^2(\{\lambda_t\}) \leq \max\!\bigl[\Omega_{\text{MLWF}}(\lambda_t)\bigr]\equiv\sigma(\{\lambda_t\}).
\end{align}
Here, $\xi_{\text{QM,max}}$ denotes the largest possible value of the quantum metric length that can be obtained from Bloch states compatible with the given CLS overlaps $\{\lambda_t\}$. The quantity $\Omega_{\text{MLWF}}$ is the spread (variance) of a maximally localized Wannier function (MLWF)~\cite{maxwf}, and $\sigma(\{\lambda_t\})$ denotes the maximum possible MLWF spread for that set of overlaps. Equation~\eqref{eqmmax} states that the quantum metric length cannot exceed the maximal Wannier spread allowed by the CLS data. In this way, bounding $\sigma(\{\lambda_t\})$ yields a bound on $\eqm$.

We now illustrate this construction explicitly in the simplest case, where the flat band has only nearest-neighbor CLS overlaps $\lambda_{\pm1}$ and these overlaps are real. Consider a CLS with support on two adjacent unit cells, which for concreteness we take to consist of three orbitals as in Eq.~\eqref{eq c}:
\begin{align}\label{eq c}
|\text{CLS};R\rangle
= a|R,\alpha\rangle
+ b|R+1,\alpha\rangle
+ c|R,\beta\rangle,
\qquad
|a|^2+|b|^2+|c|^2=1,
\end{align}
where $|R,\alpha\rangle$ denotes orbital $\alpha$ in unit cell $R$. In this section we set the lattice constant to unity. Using the gauge choice of Eq.~\eqref{eq clseig2}, the corresponding Bloch vector can be written
\begin{align}\label{bls}
u(k)
= \frac{1}{\sqrt{|a|^2 + |b|^2 + |c|^2 + a^*b e^{-ik}+ab^* e^{ik}}} 
\begin{bmatrix}
a + b e^{-i k} \\
c
\end{bmatrix}
= f(k)\,v(k),
\end{align}
where
\begin{align}
f(k)=\frac{1}{\sqrt{1+2\lambda_1 \cos(k)}},\quad
v(k)=
\begin{bmatrix}
a + b e^{-i k} \\
c
\end{bmatrix},
\qquad
\lambda_1=a^*b=ab^*.
\end{align}
The decomposition $u(k)=f(k)v(k)$ is chosen so that $f(k)$ contains only the CLS-overlap data: in particular, $f(k)=1/\sqrt{\sum_t \lambda_t e^{ikt}}$. The remaining factor $v(k)$ carries the detailed orbital weights of the CLS.

The Wannier function can then be expressed as a convolution of the Fourier transforms of $f$ and $v$. If $n$ labels the unit cell index, the $i^\text{th}$ component of the Wannier function is
\begin{align}\label{eq wann}
W_i(n)
= \int \frac{dk}{2\pi} f(k)\,v_i(k)\,e^{ikn}
= \sum_{n'} F(n')\,V_i(n-n'),
\end{align}
where
\begin{align}
F(n)=\int \frac{dk}{2\pi} f(k)e^{ikn},
\qquad
V_i(n)=\int \frac{dk}{2\pi} v_i(k)e^{ikn}.
\end{align}
Here $v(k)$ is a two-component vector, so $W_i(n)$ is obtained componentwise. The key point is that $v(k)$ Fourier-transforms to a finite set of delta functions (reflecting the fact that the CLS lives on at most two neighboring unit cells), while $f(k)$ is entirely determined by $\lambda_1$. As a result, the overall spatial variance of the Wannier function is controlled by $\lambda_1$, with only subleading dependence on the detailed amplitudes $a,b,c$.

For the concrete example in Eq.~\eqref{bls}, we obtain
\begin{align}
W_1(n)
&=\frac{1}{2\pi N}\int dk\,\frac{a+be^{-ik}}{\sqrt{1+2\lambda_1 \cos(k)}}e^{ikn}
= \sum_{n'=-\infty}^{\infty} F(n')
\{a\,\delta_{n,n'}+b\,\delta_{n,n'+1}\},\\
W_2(n)
&=\frac{1}{2\pi N}\int dk\,\frac{c}{\sqrt{1+2\lambda_1 \cos(k)}}e^{ikn}
= \sum_{n'=-\infty}^{\infty} F(n')\,c\,\delta_{n,n'}.
\end{align}

To parametrize how the CLS weight is distributed between the two unit cells, we define
\begin{align}
t=\sum_{\alpha} c_{0,\alpha}^2
\quad\text{for}\quad
|\text{CLS};0\rangle
=\sum_{\alpha} c_{0,\alpha}|0,\alpha\rangle
+ c_{1,\alpha}|1,\alpha\rangle,
\end{align}
which takes values in $(0,1)$. The parameter $t$ measures how much of the CLS weight sits in the first of the two unit cells (relative to the second). Thus $t=1/2$ corresponds to a CLS that is symmetrically shared between the two neighboring cells, whereas $t\neq 1/2$ indicates a bias toward one cell. Geometrically, when the weight is symmetric ($t=1/2$), the Wannier center sits midway between the two cells, which maximizes the spatial spread; for $t$ strongly different from $1/2$, the state is more localized around one cell and its variance is smaller.

Using Parseval's theorem, $\langle W|n^2|W \rangle$ and $\langle W|n|W \rangle$ can be written as
\begin{align}\label{eq n2}
\langle W|n^2|W \rangle &= \sum_{n=-\infty}^\infty
\left[
n^2 F(n)^2
+ t F(n)^2
+ 2\lambda_1 n^2 F(n)F(n-1)
\right]\\
&= \int_{-\pi}^\pi \frac{dk}{2\pi} \Bigg\{
-\frac{1}{\sqrt{1 + 2\lambda_1 \cos(k)}} \frac{d^2}{dk^2} \left( \frac{1}{\sqrt{1 + 2\lambda_1 \cos(k)}} \right)
+ \frac{t}{1 + 2\lambda_1 \cos(k)}\nonumber\\
&\quad- \frac{2\lambda_1 \cos(k)}{\sqrt{1 + 2\lambda_1 \cos(k)}} \frac{d^2}{dk^2} \left( \frac{1}{\sqrt{1 + 2\lambda_1 \cos(k)}} \right)
\Bigg\},\nonumber
\end{align}
\begin{align}\label{eq n2-2}
\langle W|n|W \rangle &= \sum_{n=-\infty}^\infty
\left[
t F(n)^2
+ 2\lambda_1 n F(n)F(n-1)
\right]\\
&= \int_{-\pi}^\pi \frac{dk}{2\pi} \left\{
\frac{t}{1 + 2\lambda_1 \cos(k)}
- \frac{2\lambda_1 \sin(k)}{\sqrt{1 + 2\lambda_1 \cos(k)}} \frac{d^2}{dk^2} \left( \frac{1}{\sqrt{1 + 2\lambda_1 \cos(k)}} \right)
\right\}\nonumber\,.
\end{align}
For a given $\lambda_1$, the Wannier variance $\langle n^2\rangle - \langle n\rangle^2$ is maximized at $t=1/2$, yielding
\begin{align}\label{eq xim}
\sigma_1(\lambda_1)
= \frac{1}{4\sqrt{1-4\lambda_1^2}},
\end{align}
which is a monotonically increasing function of $|\lambda_1|$.

We now show that $\sigma_1(\lambda_1)$ in Eq.~\eqref{eq xim} indeed gives the largest possible MLWF spread for fixed $\lambda_1$. Specifically, we claim that the MLWF with maximal variance,
\[
\max\!\bigl[\Omega_{\text{MLWF}}(\lambda_1)\bigr],
\]
is obtained by Fourier transforming Bloch states of the form
\begin{align}\label{eq u con}
    u_{\alpha}(k)=
    \begin{cases}
        c_{\alpha,+}\,(1+a\,e^{-ik})\\[2pt]
        c_{\alpha,-}\,(a+e^{-ik})
    \end{cases}
\end{align}
subject to
\begin{align}\label{eq u con2}
    \sum_{\alpha}c_{\alpha,+}^2
    =
    \sum_{\alpha}c_{\alpha,-}^2.
\end{align}
This ansatz is a superposition of amplitudes on two neighboring unit cells with a relative phase $e^{-ik}$. The constraint in Eq.~\eqref{eq u con2} enforces a balanced distribution between those two cells.

The logic is as follows. A maximally localized Wannier function $|W\rangle$ is an eigenstate of $PxP$~\cite{maxwf},
\begin{align}\label{eq pxp}
PxP|W\rangle =\gamma|W\rangle,
\end{align}
which implies the parallel-transport condition
\begin{align}\label{maxcon}
\frac{\partial}{\partial k}
\Bigl[
u(k)^{\dagger}\frac{\partial u(k)}{\partial k}
\Bigr]
=0.
\end{align}
The Bloch vectors in Eq.~\eqref{eq u con}, together with Eq.~\eqref{eq u con2}, satisfy Eq.~\eqref{maxcon}, hence their Fourier transforms produce MLWFs. For any other Bloch eigenvector $u'(k)$ with the same $|\lambda_1| = \tfrac{a}{1+a^2}$ but not of the form in Eq.~\eqref{eq u con}, the corresponding Wannier function $W'$ cannot both (i) have $t=1/2$ \textit{and} (ii) satisfy Eq.~\eqref{eq pxp}. Either its weight is off-centered ($t\neq 1/2$), which reduces the variance below $\sigma_1(\lambda_1)$. We thus conclude that $\sigma_1(\lambda_1)$ in Eq.~\eqref{eq xim} is indeed the maximum possible MLWF variance for a given $\lambda_1$.

Since both $\ef(\lambda_1)$ and $\sigma_1(\lambda_1)$ increase monotonically with $|\lambda_1|$, there is a one-to-one correspondence between $\ef$ and $\sigma_1$ in this nearest-neighbor case. From $\ef(\lambda_1)$ we obtain
\[
|\lambda_1| = \frac{1}{2\cosh(1/\ef)},
\]
and hence
\begin{align}\label{maxcon1}
\eqm^2(|\lambda_1|)
\leq
\sigma_1(|\lambda_1|)
=
\sigma_1\!\left(\frac{1}{2\cosh(1/\ef)}\right)
=
\frac{1}{4\sqrt{1-\tfrac{1}{\cosh^2(1/\ef)}}}.
\end{align}
Equation~\eqref{maxcon1} therefore provides the explicit upper bound $f_1(\ef)$ on $\eqm^2$ for flat bands whose CLS overlaps are restricted to nearest neighbors.

\begin{figure}
\centering
\begin{overpic}[width=0.4\linewidth]{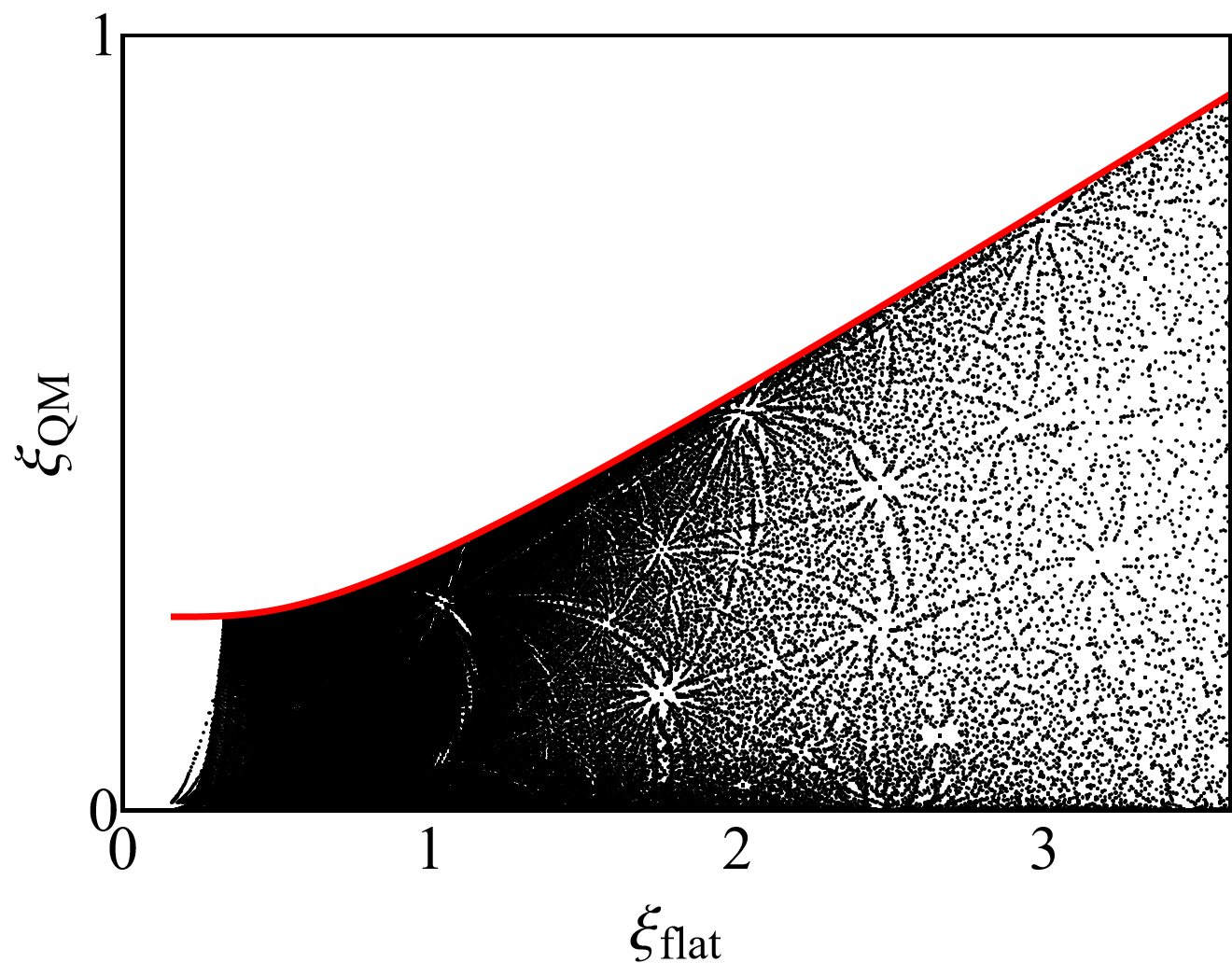}
\end{overpic}
\caption{\label{fig:scatter}Scatter plot of $(\ef,\eqm)$ for multiple CLS models. Each CLS consists of two unit cells and two intracell orbitals and has the form $|\text{CLS};R\rangle=\frac{1}{N}\big(|R,\alpha\rangle+a|R,\beta\rangle+ b|R+1,\alpha\rangle+c|R+1,\beta\rangle\big)$, with normalization factor $N$. The coefficients $a$, $b$, and $c$ were varied over $(-10,10)$ in steps of $0.2$, and $\ef,\eqm$ were computed for each choice. The points saturate the analytic bound, confirming that the function $f_1$ indeed captures the optimal (tight) relation between $\ef$ and $\eqm$ in the nearest-neighbor case.}
\end{figure}

\subsection{Nonzero $\lambda_t$ for $t\geq2$}

We now generalize to flat bands whose CLS overlaps extend beyond nearest neighbors. Suppose $\lambda_t$ may be nonzero for $|t|>1$, so that a single CLS has support on three or more unit cells. In this case, the procedure for obtaining the bound function $f_{\tmx}$ is almost the same as above, with one important difference: $\sigma(\{\lambda_t\})$ and $\ef(\{\lambda_t\})$ are no longer in strict one-to-one correspondence.

The inequality $\eqm^2 \leq \sigma(\{\lambda_t\})$ from Eq.~\eqref{eqmmax} still holds. Thus the problem becomes: for a fixed $\ef$, what choice of CLS overlaps $\{\lambda_t\}$ maximizes $\sigma(\{\lambda_t\})$, i.e., gives the largest possible Wannier spread? Intuitively, the Wannier function is a convolution of the CLS coefficients with the envelope $F(n)$ [cf.~Eq.~\eqref{eq wann}]. To maximize its variance at a given $\ef$, one should push the CLS weight as far apart as possible while keeping only two dominant peaks. This suggests that the extremal case is realized when the CLS occupies only two unit cells separated by $\tmx$, so that only $\lambda_0$ and $\pm\lambda_{\tmx}$ are nonzero and all other $\lambda_t$ vanish. Indeed, for fixed $\ef$, the MLWF variance increases with $T$ under this restriction.

Based on this assumption (rigorously proven for $\tmx=1$, and expected to be conservative for $\tmx>1$), we proceed by keeping only $\pm\lambda_{\tmx}$ nonzero. Generalizing Eqs.~\eqref{eq n2}–\eqref{eq n2-2}, we obtain
\begin{align}\label{eq gh2}
\sigma_{\tmx}(\lambda_{\tmx})
&=\sum_{n=-\infty}^\infty \bigg[
n^2 F(n)^2
+ \Bigl(\frac{\tmx}{2}\Bigr)^2 F(n)^2
+ 2\lambda_{\tmx} \Bigl(n-\frac{\tmx}{2}\Bigr)^2 F(n)F(n-\tmx)
\bigg]\\
&= \int_{-\pi}^\pi \frac{dk}{2\pi} \Bigg[
-\frac{1}{\sqrt{1 + 2\lambda_{\tmx} \cos(\tmx k)}} \frac{d^2}{dk^2} \!\left( \frac{1}{\sqrt{1 + 2\lambda_{\tmx} \cos(\tmx k)}} \right)
+\Bigl(\frac{\tmx}{2}\Bigr)^2\frac{1}{1 + 2\lambda_{\tmx} \cos(\tmx k)}\nonumber\\
&\hspace{5.8em}
- \frac{2\lambda_{\tmx} e^{-i\tmx k/2}}{\sqrt{1 + 2\lambda_{\tmx} \cos(\tmx k)}} \frac{d^2}{dk^2} \!\left( \frac{e^{-i\tmx k/2}}{\sqrt{1 + 2\lambda_{\tmx} \cos(\tmx k)}} \right)
\Bigg]\nonumber\\
&=\frac{\tmx^2}{4\sqrt{1-4\lambda_{\tmx}^2}}.\nonumber
\end{align}
This $\sigma_{\tmx}(\lambda_{\tmx})$ is the maximal MLWF variance achievable when the CLS has only two peaks separated by $\tmx$ unit cells.

The Wannier function that realizes this variance is obtained, as before, by Fourier transforming Bloch eigenvectors of the form
\begin{align}\label{eq ucon}
    u_{\alpha}(k)=
    \begin{cases}
        c_{\alpha,+}\,(1+a\,e^{-i\tmx k})\\[2pt]
        c_{\alpha,-}\,(a+e^{-i\tmx k})
    \end{cases}
\end{align}
subject to
\begin{align}\label{eq ucon2}
    \sum_{\alpha}c_{\alpha,+}^2
    =
    \sum_{\alpha}c_{\alpha,-}^2.
\end{align}
This is again the most general two-site CLS in momentum space, now with the two sites separated by $\tmx$ unit cells rather than one. The balanced condition in Eq.~\eqref{eq ucon2} enforces $t=\tmx/2$, i.e., equal weight on the two cells.

\begin{figure}
\centering
\begin{overpic}[width=0.3\linewidth]{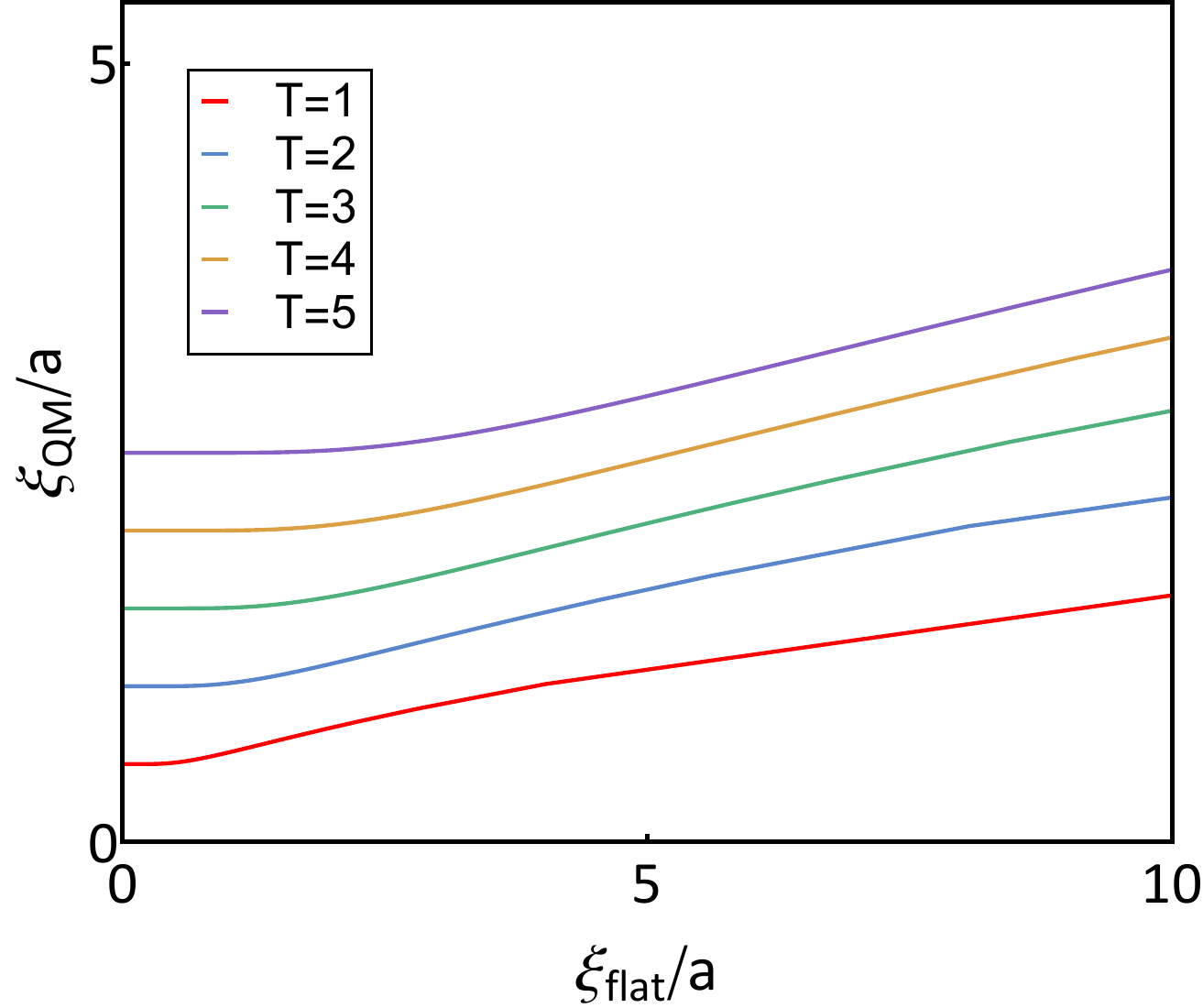}
\end{overpic}
\begin{overpic}[width=0.3\linewidth]{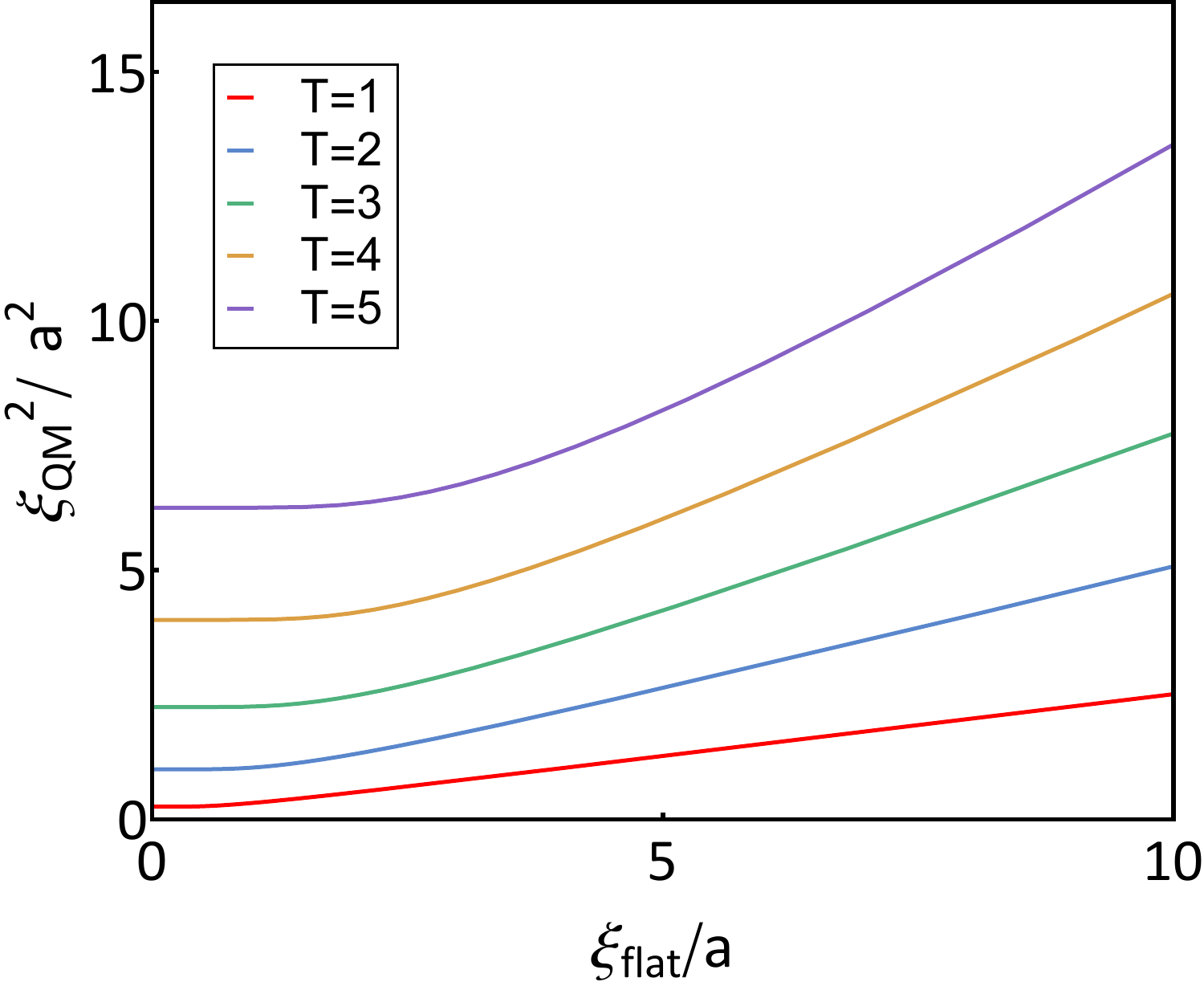}
\end{overpic}
\caption{\label{fig:qmbound} (a) Bound function describing the relation between \(\ef\) and \(\eqm\) for several values of \(t_{\text{max}}\). 
(b) When \(\ef\) is plotted against \(\eqm^{2}\), the bound function exhibits an approximately linear behavior in the large-\(\ef\) regime.
}
\end{figure}

When only $\pm\lambda_{\tmx}$ are nonzero, the flat-band length $\ef$ is
\begin{align}
    \ef
    \;=\;
    \tmx\Bigl[\,
      \operatorname{arccosh}\!\Bigl(\frac{1}{|2\lambda_{\tmx}|}\Bigr)
    \Bigr]^{-1}.
\end{align}
Since both $\sigma_{\tmx}(\lambda_{\tmx})$ and $\ef$ are monotonically increasing functions of $|\lambda_{\tmx}|$, we obtain
\begin{align}\label{maxcon11}
\eqm^2\;\leq\;\sigma_{\tmx}(\lambda_{\tmx})
=\sigma_{\tmx}\!\left(\frac{1}{2\cosh(\tmx/\ef)}\right)
=\frac{\tmx^2}{4\sqrt{\,1-\frac{1}{\cosh^2(\tmx/\ef)}\,}}.
\end{align}
Equation~\eqref{maxcon11} is the desired upper bound $f_{\tmx}(\ef)$ for general $\tmx$, i.e., for CLSs whose nonzero overlaps extend across $\tmx$ unit cells. The corresponding bound function is shown in Fig.~\ref{fig:qmbound}.

The coefficient $\lambda_{\tmx}$ is constrained by $|\lambda_{\tmx}|<\tfrac{1}{2}$. In the limit $|\lambda_{\tmx}|\to \tfrac{1}{2}-$, $\ef$ diverges. In this regime, we find the slope ratio
\begin{align}\label{eq:slope-ratio}
\lim_{|\lambda_{\tmx}|\rightarrow\frac{1}{2}-}
\frac{\partial \ef/\partial |\lambda_{\tmx}|}{\partial \sigma_{\tmx}/\partial |\lambda_{\tmx}|}
=\frac{4}{\tmx}.
\end{align}
Thus, in the large-$\ef$ limit,
\begin{align}
  \ef
  \;\gtrsim\;
  \frac{4}{\tmx}\,\eqm^{2}.
\end{align}
This is the asymptotic form quoted in the main text: for very extended flat bands (large $\ef$), $\ef$ is bounded from below by a quantity proportional to $\eqm^2$, with a prefactor determined only by the maximum CLS separation $\tmx$.

\section{Application to nearly flat bands}

\subsection{Effective localization length for a nearly flat band}

In the main text, we defined the flat-band length scale $\ef$ from the large-distance decay of the Fourier transform of the flat-band Bloch vector. For an exactly flat band with Bloch vector \(u_\alpha(k)\), we introduced
\[
F_\alpha(r)
=
\int_{-\pi}^{\pi} \frac{dk}{2\pi}\, e^{ikr}\,
u_\alpha(k)\,u_\alpha^{*}(k),
\]
and defined $\ef$ as the localization length characterizing the asymptotic decay
\[
F_\alpha(r) \propto e^{-|r|/\ef}
\quad (|r|\to\infty).
\]
Because the compact localized state (CLS) structure fixes the analytic properties of \(u_\alpha(k)\,u_\alpha^{*}(k)\) in complex momentum, $\ef$ is independent of the choice of orbital \(\alpha\).

We now extend this construction to a nearly flat band with normalized Bloch vector \(u_{\mathrm{n.f.},\alpha}(k)\). For each orbital component \(\alpha\), we define
\begin{align}
F_{\mathrm{n.f.},\alpha}(r)
=
\int_{-\pi}^{\pi} \frac{dk}{2\pi}\, e^{ikr}\,
u_{\mathrm{n.f.},\alpha}(k)\,u_{\mathrm{n.f.},\alpha}^{*}(k),
\end{align}
and extract an effective localization length \(\xi_{\mathrm{eff},\alpha}\) from the large-distance behavior
\begin{align}
F_{\mathrm{n.f.},\alpha}(r) \propto
e^{-|r|/\xi_{\mathrm{eff},\alpha}}
\quad (|r|\to\infty).
\end{align}
Unlike the exactly flat band case, \(\xi_{\mathrm{eff},\alpha}\) is in general orbital dependent. However, when the band dispersion is sufficiently small and the nearly flat band remains well isolated, the nearly flat band can be regarded as perturbed flat band and \(\xi_{\mathrm{eff},\alpha}\) for different \(\alpha\) are numerically very close to one another. In practice, we therefore use a representative effective localization length \(\xi_{\mathrm{eff}}\) (for example, obtained from a specific orbital or from an average over \(\alpha\)), noting that the orbital dependence is negligible in the nearly flat regime.

In this section, we develop a perturbative scheme to approximate \(\xi_{\mathrm{eff}}\) starting from an ideal flat-band model. We consider a nearly flat band obtained by adding a weak perturbation \(V\) to a perfect flat band Hamiltonian,
\begin{align}
H_{\mathrm{n.f.}} = H_{\mathrm{flat}} + \lambda V,
\quad |\lambda|\ll 1,
\end{align}
and denote the corresponding Bloch vectors by \(u_{\mathrm{n.f.},\alpha}(k)\) and \(u_\alpha(k)\), respectively. For small \(|\lambda|\), the change in the Bloch vector can be written schematically as
\begin{align}
u_{\mathrm{n.f.},\alpha}(k)
=
u_\alpha(k) + \delta u_\alpha(k),
\end{align}
with \(\delta u_\alpha(k)\) controlled by the weak dispersion. Provided that the perturbation does not close the gaps to neighboring bands and does not introduce new singularities in the complex-\(k\) plane, the analytic region of $u_{\mathrm{n.f.},\alpha}(k)\,u_{\mathrm{n.f.},\alpha}^{*}(k)$
coincides with that of \(u_\alpha(k)\,u_\alpha^{*}(k)\) up to corrections of order \(|\lambda|\). By the Paley–Wiener theorem, this implies that the corresponding localization lengths satisfy
\[
\xi_{\mathrm{eff},\alpha}
=
\ef + \mathcal{O}(|\lambda|),
\]
so that the effective localization length of the nearly flat band is approximately given by the flat-band length $\ef$ of the associated ideal flat-band model. 

In the Stub lattice model with a weak flatness-breaking term, we numerically confirm that \(\xi_{\mathrm{eff},\alpha}\) is nearly orbital independent and closely follows $\ef$ when the dispersion of the nearly flat band is small. In the orbital basis $(A,B,C)$, the Bloch Hamiltonian is
\begin{align}\label{nearlyh}
H(k)=
\begin{pmatrix}
2 t_1 \cos k & 1 + e^{ik} & 0 \\
1 + e^{-ik} & 0 & d \\
0 & d & 0
\end{pmatrix},
\end{align}
with $d = 0.5$. The parameter $t_1$ plays the role of a flatness-breaking perturbation: for $t_1 = 0$, the middle band is perfectly flat, while $t_1 \neq 0$ induces a small dispersion and produces a nearly flat band.

For the perfectly flat case $t_1 = 0$, we compute the flat-band length $\ef$ from the flat-band Bloch vector and obtain
\[
\ef \simeq 2.02.
\]
When a weak dispersion is introduced by choosing $t_1 = 0.01$, we evaluate the effective localization length $\xi_{\mathrm{eff},\alpha}$ from the large-distance decay of $F_{\mathrm{n.f.},\alpha}(r)$ for each orbital $\alpha = A,B,C$ and find
\[
\xi_{\mathrm{eff},A} \simeq 2.22,\quad
\xi_{\mathrm{eff},B} \simeq 2.28,\quad
\xi_{\mathrm{eff},C} \simeq 2.22.
\]
For a larger but still small perturbation $t = 0.02$, we obtain
\[
\xi_{\mathrm{eff},A} \simeq 2.42,\quad
\xi_{\mathrm{eff},B} \simeq 2.43,\quad
\xi_{\mathrm{eff},C} \simeq 2.41.
\]
These results demonstrate that (i) the orbital dependence of $\xi_{\mathrm{eff},\alpha}$ remains weak in the nearly flat regime, and (ii) the deviation from the ideal flat-band value $\ef$ grows smoothly with the flatness-breaking parameter $t_1$, consistent with the perturbative expectation
\[
\xi_{\mathrm{eff},\alpha} - \ef = \mathcal{O}(|\lambda|)\propto \mathcal{O}(|t_1|).
\]
In particular, for small $t_1$ the effective localization length of the nearly flat band remains close to the flat-band length $\ef$ of the corresponding perfect flat-band model, in agreement with the analytic argument based on the stability of the analytic region of the Bloch vector.

\subsection{Coherence length in a nearly flat band}

We now analyze the superconducting coherence length in the regime
\[
w_{\mathrm{n.f.}}\ll \Delta \ll W_{\mathrm{gap}},
\]
where \(w_{\mathrm{n.f.}}\) is the bandwidth of the nearly flat band, \(W_{\mathrm{gap}}\) is the separation from other bands, and \(\Delta\) is the superconducting gap.  
This hierarchy ensures that the band projection remains valid and interband pairing are suppressed by the large energy denominator.

When the chemical potential lies inside the nearly flat band, the anomalous correlator of orbital \(\alpha\) becomes
\begin{align}\label{eq:nf-proj}
K_{\alpha}(r)
= \frac{1}{N}\sum_{k} e^{ikr}\,
u_{\mathrm{n.f.},\alpha}(k)\,
u_{\mathrm{n.f.},\alpha}(-k)\,
\big\langle c_{k\uparrow}c_{-k\downarrow}\big\rangle.
\end{align}

The pairing amplitude is
\begin{align}\label{eq:nf-amp}
\big\langle c_{k\uparrow}c_{-k\downarrow}\big\rangle
=\frac{1}{\beta}\sum_{\omega_m}
\frac{\Delta}{
\omega_m^2+
\big(\epsilon_{\mathrm{n.f.}}(k)-\mu\big)^2+
\Delta^2 }.
\end{align}
Since \(|\epsilon_{\mathrm{n.f.}}(k)-\mu|\le w_{\mathrm{n.f.}}\ll \Delta\), the denominator varies only by relative corrections of order \((w_{\mathrm{n.f.}}/\Delta)^2\).  
Hence the pairing amplitude is \(k\)-independent to leading order.

Equation~\eqref{eq:nf-proj} is the Fourier transform of the product
\[
u_{\mathrm{n.f.},\alpha}(k)\,u_{\mathrm{n.f.},\alpha}(-k)\times 
\big\langle c_{k\uparrow}c_{-k\downarrow}\big\rangle.
\]
Let \(\gamma_u\) denote the half-width of the analytic region for  
\(u_{\mathrm{n.f.},\alpha}(k)u_{\mathrm{n.f.},\alpha}(-k)\),  
and \(\gamma_F\) that of the pairing amplitude.  
By Paley–Wiener and the fact that the Fourier transform of a product corresponds to a convolution in real space, the asymptotic decay is set by the narrower analytic strip:
\[
K_\alpha(r)\sim e^{-\min(\gamma_u,\gamma_F)|r|}
\quad\Longrightarrow\quad
\ecoh=\max(\xi_{\mathrm{eff}},\xi_{F}),
\]
where \(\xi_{\mathrm{eff}}=1/\gamma_u\) and \(\xi_F=1/\gamma_F\).

Since the pairing amplitude is $k$-independent to leading order, \(\xi_{\mathrm{eff}}\) is typically much larger than $\xi_F$ and determined by the analytic structure of the Bloch eigenstates as in flat band case.  
Thus, in the regime \(w_{\mathrm{n.f.}}\ll \Delta \ll W_{\mathrm{gap}}\),
\[
\ecoh \simeq \xi_{\mathrm{eff}},
\]
so the coherence length in a nearly flat band is controlled by the same analytic-width mechanism as in the perfectly flat case.

\begin{figure}
\centering
\begin{overpic}[width=0.4\linewidth]{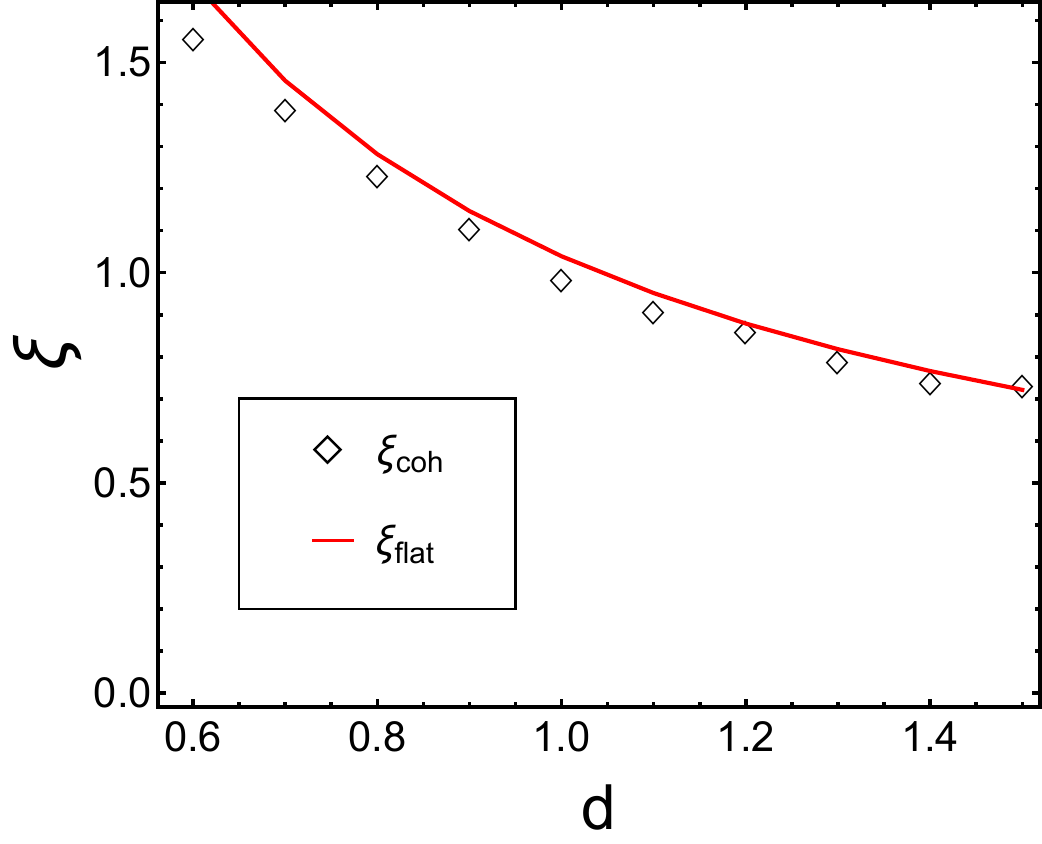}
\end{overpic}
\caption{\label{fig:nearly} Numerical coherence length in the nearly flat band Stub lattice model.
In the BdG Hamiltonian with a weak flatness-breaking term, \(|K_A(r)|\) is fitted for various values of \(d \in [0.6,1.4]\) to extract \(\xi_{\mathrm{coh}}\), which is then compared with the flat-band length \(\ef\) of the corresponding perfectly flat band.}
\end{figure}

To substantiate the above analytic argument, we now present numerical results for a nearly flat band in the same Stub lattice model. We consider the BdG Hamiltonian of the main text with the normal-state part given by Eq.~\eqref{nearlyh}. The flatness-breaking parameter is fixed to a small value \(t_1 = 2\times 10^{-4}\) and the Hubbard interaction strength is set to \(U=-0.05\). For each value of $d$, the middle band realizes a nearly flat band that remains well separated from the other bands. We place the chemical potential $\mu$ inside this nearly flat band and calculate the corresponding BdG Hamiltonian to obtain the coherence length.

From the BdG spectrum and eigenvectors, we first compute the anomalous correlation function \(K_\alpha(r)\) for each orbital \(\alpha\). Because the flatness-breaking term is very small, the differences in the effective localization length \(\xi_{\mathrm{eff},\alpha}\) between orbitals are negligible, and we therefore focus on orbital \(A\) and calculate the corresponding correlator \(K_A(r)\). The large-distance envelope of \(|K_A(r)|\) is then fitted to an exponential form
\[
|K_A(r)| \propto e^{-|r|/\xi_{\mathrm{coh}}}
\]
to extract \(\ecoh\). Since \(t_1\) is extremely small, the effective localization length \(\xi_{\mathrm{eff}}\) of the nearly flat band is expected to be very close to \(\ef\) of the corresponding ideal flat-band model at \(t_1=0\), so the fitted coherence length \(\ecoh\) can be directly compared with \(\ef\).

In Fig.~\ref{fig:nearly}, we plot the fitted coherence length \(\xi_{\mathrm{coh}}\) as a function of \(d\) and compare it with the flat-band length \(\ef\) obtained from the flat-band Bloch vectors at \(t_1=0\). Over the parameter range studied, \(\xi_{\mathrm{coh}}\) closely tracks \(\ef\), with only small deviations. This numerical behavior is fully consistent with the analytic expectation
\[
\ecoh \simeq \xi_{\mathrm{eff}} \simeq \ef
\]
in the regime \(w_{\mathrm{n.f.}}\ll \Delta \ll W_{\mathrm{gap}}\).

\subsection{Relation to the quantum metric length}

Because the nearly flat band is obtained by a perturbative deformation of a perfectly flat band, the Bloch vector \(u_{\mathrm{n.f.}}(k)\), and hence its quantum metric, vary only weakly.  
Consequently, the bound relating the quantum-metric length and the flat-band length,
\[
\eqm^2 \;\le\; f_T(\ef),
\]
remains valid up to small corrections when \(\ef\) is replaced by \(\xi_{\mathrm{eff}}\),
\[
\eqm^2 \;\lesssim\; f_T\big(\xi_{\mathrm{eff}}\big).
\]
Thus, as long as \(u_{\mathrm{n.f.}}(k)\approx u(k)\) and  
\(w_{\mathrm{n.f.}}\ll W_{\mathrm{gap}}\),  
the \((\xi,\,\eqm)\) relation derived for exactly flat bands carries over to nearly flat bands with negligible quantitative modification.

\end{document}